\documentclass[%
reprint,
superscriptaddress,
amsmath,amssymb,
mathtools,
aps, 
pra,
]{revtex4-2}

\usepackage{graphicx}
\usepackage[colorlinks=true,allcolors=blue]{hyperref}

\usepackage{diffcoeff}
\usepackage{siunitx}
\usepackage{array}

\newcommand{\SIi}{Supplementary Information}
\newcommand{\Qi}{$Q_\text{int}$}
\newcommand{\Qd}{$Q_\text{D}$}
\newcommand{\Qm}{$Q_\text{m}$}
\newcommand{\sr}{$\sigma_\text{released}$}
\newcommand{\er}{$\epsilon_\text{released}$}
\newcommand{\sg}{$\sigma_\text{residual}$}

\newcommand{\fm}{$f_\text{m}$}
\newcommand{\Qf}{$Q_\text{m}\times f_\text{m}$}

\newcommand{\eref}[1]{Eq.~\ref{#1}}
\newcommand{\fref}[1]{Fig.~\ref{#1}}
\newcommand{\tref}[1]{Tab.~\ref{#1}}

\begin{document}
	
	\title{Nanomechanical crystalline AlN resonators with high quality factors for quantum optoelectromechanics}
	
	\author{Anastasiia Ciers}
	\email{anastasiia.ciers@chalmers.se}
	\author{Alexander Jung}
	\author{Joachim Ciers}
	\author{Laurentius Radit Nindito}
	\author{Hannes Pfeifer}
	\affiliation{Department of Microtechnology and Nanoscience (MC2),\\
		Chalmers University of Technology, SE-412 96 Gothenburg, Sweden}
	\author{Armin Dadgar}
	\author{Andr\'e Strittmatter}
	\affiliation{Institute of Physics, Otto-von-Guericke-University Magdeburg, 39106 Magdeburg, Germany
	}%
	\author{Witlef Wieczorek}%
	\email{witlef.wieczorek@chalmers.se}
	\affiliation{Department of Microtechnology and Nanoscience (MC2),\\
		Chalmers University of Technology, SE-412 96 Gothenburg, Sweden}
	
	\begin{abstract}
		High-\Qm{} mechanical resonators are crucial for applications where low noise and long coherence time are required, as mirror suspensions, quantum cavity optomechanical devices, or nanomechanical sensors. Tensile strain in the material enables the use of dissipation dilution and strain engineering techniques, which increase the mechanical quality factor.
		These techniques have been employed for high-\Qm{} mechanical resonators made from amorphous materials and, recently, from crystalline materials such as InGaP, SiC, and Si. A strained crystalline film exhibiting substantial piezoelectricity expands the capability of high-\Qm{} nanomechanical resonators to directly utilize electronic degrees of freedom. In this work we realize nanomechanical resonators with \Qm{} up to $2.9\times 10^{7}$ made from tensile-strained \SI{290}{\nano\meter}-thick AlN, which is an epitaxially-grown crystalline material offering strong piezoelectricity. 
		We demonstrate nanomechanical resonators that exploit dissipation dilution and strain engineering to reach a \Qf-product approaching
		$10^{13}$\,\SI{}{\hertz} at room temperature. We realize a novel resonator geometry, triangline, whose shape follows the Al-N bonds and offers a central pad that we pattern with a photonic crystal. This allows us to reach an optical reflectivity above 80\% for efficient coupling to out-of-plane light. The presented results pave the way for quantum optoelectromechanical devices at room temperature based on tensile-strained AlN.
	\end{abstract}

	\maketitle
	
	\section{Introduction}
	Engineering of tensile-strained materials has enabled rapid progress in realizing nanomechanical resonators with ever-higher quality factors \cite{sementilli2022nanomechanical}. The low mass and high quality factor of a nanomechanical resonator result in low thermal force noise, which enables measuring small forces as, e.g., required for the detection of single-proton spins \cite{eichler2022ultra} or gravity between small masses \cite{westphal2021measurement}. Furthermore, a high-\Qf{}-product increases the number of coherent oscillations, which is essential for realizing quantum opto- or electromechanical devices \cite{midolo2018nano} for the use in quantum technologies \cite{barzanjeh2022optomechanics}.
	
	Nanomechanical resonators with record-high quality factors of $10^{10}$ \cite{cupertino2024centimeter} have been predominantly achieved in tensile-strained amorphous Si$_3$N$_4$, employing dissipation dilution, soft clamping, and strain engineering techniques \cite{unterreithmeierDampingNanomechanicalResonators2010,yu2012control,tsaturyan2017ultracoherent,ghadimi2018elastic,fedorov2019generalized}. These techniques have led to a variety of nanomechanical resonator geometries, including 2D phononic crystal (PnC)-shielded membranes \cite{tsaturyan2017ultracoherent}, 1D PnC beams \cite{ghadimi2018elastic}, hierarchically-clamped devices \cite{bereyhi2022hierarchical}, and resonators optimized through machine learning methods \cite{shin2022spiderweb,hoj2021ultra}. Recently, tensile-strained crystalline materials made from InGaP \cite{cole2014tensile,buckle2018stress,manjeshwar2023high}, SiC \cite{romero2020engineering}, or Si \cite{beccari2022strained} have been investigated for high-\Qm{} mechanical resonators. The latter work demonstrated 1D PnC beams in Si with \Qm{} of $10^{10}$ at cryogenic temperatures \cite{beccari2022strained}. Crystalline materials have fewer defects, which potentially leads to a larger intrinsic quality factor, and, thus, to an enhanced diluted quality factor. Furthermore, highly ordered materials, depending on their crystal structure, can offer additional functionality, such as electrical conductivity, piezoelectricity, or superconductivity. This would enable interfacing mechanical vibrations directly to electronic degrees of freedom \cite{cimalla2007group,brueckner2011micro,midolo2018nano} without requiring the deposition of additional materials on high-\Qm{} nanomechanical resonators \cite{brubaker2022optomechanical,seis2022ground}, which increases fabrication complexity and may decrease \Qm{} \cite{yu2012control}. 
	
	In this work, we demonstrate high-\Qm{} nanomechanical resonators made from tensile-strained AlN. This crystalline material is non-centrosymmetric, thus, pyro- and piezoelectric, and has so far been utilized in unstrained GHz mechanics \cite{o2010quantum,mirhosseini2020superconducting}. Moreover, AlN is widely used in ultraviolet photonics \cite{Morkoc2008,Kneissl2016} and can host defect centers \cite{xue2020single} that act as single-photon emitters. AlN is chemically stable \cite{taylor1960some} and provides a wide bandgap (6.2\,eV at \SI{300}{\kelvin}) with a broad transparency window that covers the ultraviolet to mid-infrared range. Hence, AlN is free from two-photon absorption at telecom wavelengths \cite{liu2023aluminum}, contrary to Si. 
	These capabilities make AlN and in general III-nitrides suitable materials for realizing a hybrid platform for interfacing electrical, mechanical and optical degrees of freedom \cite{midolo2018nano}. 
	
	We realize AlN nanomechanical resonators with \Qm{} as high as $2.9 \times 10^7$ and \Qf-product close to $10^{13}$\,\SI{}{\hertz} at room temperature. 
	Mechanical resonators in crystalline III-nitrides have been demonstrated in unstrained AlN \cite{cleland2001single,brueckner2005electromechanical,pernice2012high,ghasemi2016ultra} or compressively strained GaN \cite{sang2020strain}, but these structures could not profit from dissipation dilution, soft clamping, or strain engineering techniques. Although there are examples of strained AlN resonators \cite{placidi2009highly}, no high-\Qm{} devices have been reported. 
	In this work, we apply dissipation dilution, soft clamping, and strain engineering techniques to a \SI{290}{\nano \meter}-thick crystalline AlN film of residual tensile stress of \SI{1.4}{\giga\pascal} grown on a Si(111) substrate \cite{cimalla2007group}. We demonstrate high-\Qm{} nanomechanical resonators with different geometries: uniform beams, tapered 1D PnC beams, and hierarchically-clamped structures. We compare the experimental results with eigenfrequency simulations and dissipation dilution calculations of a prestressed crystalline material with hexagonal symmetry. We present a new resonator type, triangline, whose geometry has three-fold symmetry and follows the Al-N bonds in the crystal structure. Importantly, this hierarchically-clamped triangline provides a central pad that we pattern with a photonic crystal (PhC), which allows us to engineer the pad's out-of-plane optical reflectivity \cite{fan2002analysis,kini2020suspended,enzian2023phononically,zhou2023cavity}.
	
	\section{Results}
	
	\subsection{Fabrication}
	\begin{figure*}[ht!]
		\centering
		\includegraphics[width=\textwidth]{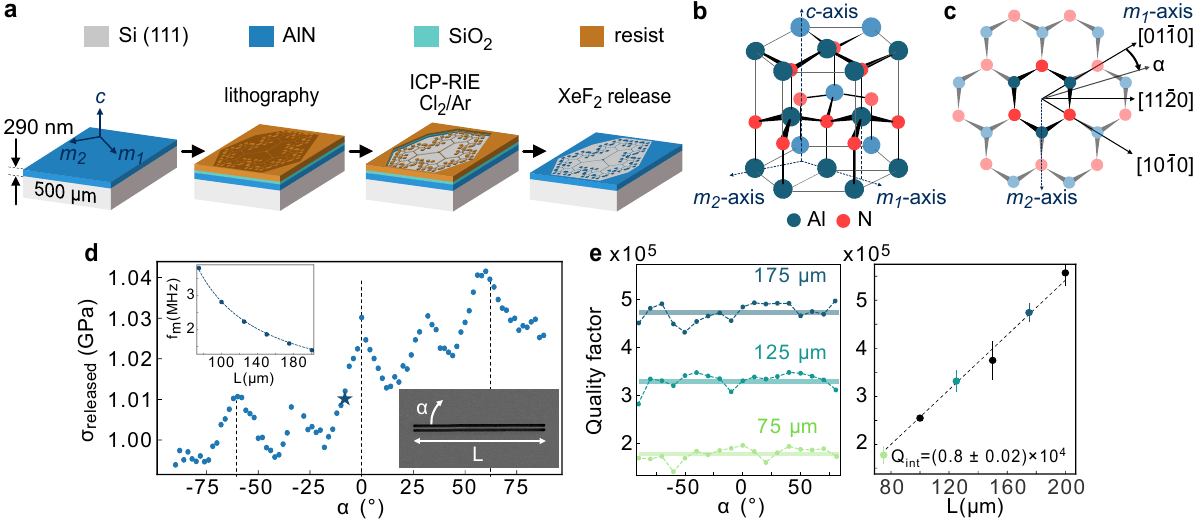}
		\caption{\textbf{Material characterization of AlN nanomechanical resonators.} a. Illustration of the fabrication process. b. 3D view of the wurtzite AlN crystal, where the $m_1$ and $m_2$ axes point along two mirror planes of the crystal, and $\alpha$ denotes the in-plane rotation angle with respect to the $m_1$-axis. c. Top view of the wurtzite AlN crystal. d. Released in-plane stress, the insets show the fit to \eref{eq:freq_beam} for the data-point indicated by the star, and an SEM image of a \SI{175}{\micro\meter}-long beam. e. The mechanical quality factor is largely independent of the in-plane orientation of the beam, $\alpha$ (left), and increases linearly with beam length, $L$ (right). The horizontal lines depict the mean of \Qm{} (left). The black dashed line is a fit of \Qi{}  (right).}
		\label{fig:material}
	\end{figure*}
	The fabrication process steps are summarized in \fref{fig:material}a (details in the Experimental Section and \SIi). 
	The \SI{290}{\nano\meter}-thick wurzite AlN is grown by metal-organic vapour-phase epitaxy (MOVPE) on a (111)-oriented 500\,$\mu$m-thick Si wafer. The roughness of the grown film is $1 \pm 0.1$\,\SI{}{\nano\meter} (root mean square, AFM data in \SIi). The resonator geometry is defined by electron-beam lithography. The exposed pattern is then transferred to the underlying AlN layer by ICP-RIE etching with a Cl$_2$/Ar mixture using a SiO$_2$ hard mask. 
	Subsequently, the resist and hard mask are removed and in a final step the AlN resonators are released in a dry release step with XeF$_2$.
	Such isotropic release allows the pattern to be independent from the substrate orientation, which is not possible with a KOH release etch of Si$_3$N$_4$ on a Si(100) wafer \cite{bereyhi2022hierarchical}, but similar to a dry release etch of amorphous SiC \cite{xu2023high} or Si$_3$N$_4$ \cite{cupertino2024centimeter} on Si.
	Furthermore, through the dry release process we achieve a high fabrication yield of above 90\%. This is contrary to a wet release (e.g., KOH-based), which requires additional lithography, etching steps and critical point drying, to increase the membrane-substrate gap and ensure high fabrication yield for large area resonators \cite{bereyhi2022hierarchical}.
	The presented fabrication process allows the realization of a range of high-\Qm{} nanomechanical resonator geometries including, but not limited to, uniform beams (inset \fref{fig:material}d), tapered 1D PnC beams (\fref{fig:1D_PnC}a, b), and hierarchically-clamped trianglines (\fref{fig:triangline}a).
	
	\subsection{Material characterization}
	Wurtzite AlN belongs to the $P_{6_3mc}$ space group with a polar axis along the [001] direction, i.e., the $c$-axis. We illustrate its crystal structure in \fref{fig:material}b and \fref{fig:material}c.
	The AlN crystal has a 19\% lattice-mismatch with the silicon substrate underneath \cite{liu2003atomic}, thus, its atoms are displaced from their equilibrium positions resulting in an approximately \SI{20}{\nano\meter}-thick defect-rich layer at the interface of the AlN film and the Si substrate \cite{mante2018proposition} (TEM images in \SIi). The quality of the film improves further away from the substrate \cite{mastro2006mocvd}, while at the same time introducing a strain gradient.
	
	The AlN film has a thickness of \SI{290}{\nano \meter} and a refractive index of 2.1 in the telecom range, determined via ellipsometry (see \SIi). In this work, we do not make use of the piezoelectric properties of the film, but have verified that our AlN film is indeed piezoelectric. We determined an effective piezoelectric coefficient $d_{33,\text{eff}}$ of $1.8$\,pm/V of the AlN film rigidly-clamped to the silicon substrate (for details see \SIi{}).
	
	We determine the strain in the AlN film by Raman measurements of the AlN $E_2^\text{high}$ phonon mode \cite{callsen2014phonon}.
	We observe $E_2^\text{high}$ at \SI{650.67}{}\,cm$^{-1}$, which corresponds to an average residual stress of the AlN film, \sg{}, of $1.43\pm0.01$\,\SI{}{\giga\pascal} as targeted in the film growth (see Methods \ref{subsec:Raman}). For simplicity, we assume that the AlN crystal exhibits hexagonal symmetry (for details see \SIi{}), which yields relations for the elastic constants as ${C_{11} = C_{22}}$, ${C_{13} = C_{23}}$ and ${C_{66} = (C_{11} - C_{12})/2}$. Deformations in the $c$-plane of the hexagonal crystal are then determined by two elastic constants only \cite{li2022tailorable}, making the model comparable to isotropic materials, like amorphous Si$_3$N$_4$.
	
	To measure the released stress and evaluate the intrinsic mechanical quality factor, \Qi{}, we pattern beams of various lengths (75 to 200\,$\mu$m) and rotation angles $\alpha$ ($-$\SI{90}{\degree} to +\SI{90}{\degree}), as illustrated in the inset of \fref{fig:material}d. 
	The AlN beams have stress-dominated mechanical frequencies, \fm{}, with the fundamental mode frequency given as \cite{klass2022determining}
	\begin{equation}
		f_\text{m} = \frac{1}{2L}\sqrt{\frac{\text{\sr}}{\rho} },
		\label{eq:freq_beam}
	\end{equation}
	where $L$ is the length of the beam and $\rho$ is the AlN density (parameters in Methods, \tref{tab:param}). We measured \fm{} and \Qm{} of the beams in high vacuum ($7 \times 10^{-6}$\,\SI{}{\milli\bar}) at room temperature using an optical interferometric position measurement setup (details in Methods \ref{subsec:Characterization}). We determined Young's modulus of the AlN film to be $E = 270\pm$\,\SI{10}{\giga\pascal} from measurements of the resonant frequencies of higher order modes of the beam \cite{klass2022determining} (see \SIi).
	The released stress of the beams extracted from measurements of their fundamental mode frequencies using \eref{eq:freq_beam} is shown in \fref{fig:material}d. The released stress is close to the expected value of $(1-\nu)$\sg{} = \SI{1}{\giga\pascal} (with $\nu=0.28$ is the Poisson ratio). Instead of a constant released stress expected from a hexagonal symmetry, we observe a weak in-plane anisotropy. We obtain a small periodic stress variation with an amplitude of about \SI{10}{\mega \pascal}.
	We attribute the \SI{60}{\degree}-periodicity to the AlN crystal structure \cite{zeng2017crystal} and
	the additional \SI{30}{\degree}-periodicity could be the result of crystal twinning \cite{parsons2003introduction} (for the discussion see \SIi).
	
	The quality factor of a strained high-aspect-ratio mechanical resonator is enhanced by the dilution factor, $D_Q$, over the intrinsic quality factor, \Qi{}, via \cite{fedorov2019generalized}
	\begin{equation}
		Q_\text{D} = D_Q \ Q_\text{int}.
		\label{eq:dissipation_dilution}
	\end{equation}
	While \Qi{} is a material property, inversely proportional to the delay between stress and strain, $D_Q$ is engineered by the resonator geometry and depends on the linear and non-linear dynamic contributions to the elastic energy of specific mode shapes (see Methods for the explicit formulae).
	
	We determine \Qi{} from measurements of $Q_\text{beam}$ of the long, thin strained beams. Their quality factor is limited by dissipation dilution and given as \cite{schmid2011damping}
	\begin{equation}
		Q_\text{beam} =  D_{Q}^{\mathrm{beam}}Q_\text{int},
		\label{eq:Qint}
	\end{equation}
	with $D_{Q}^{\mathrm{beam}}=\left[{(\pi\lambda)^2 + 2\lambda}\right]^{-1}$, the stress parameter $\lambda = \frac{h}{L} (12\text{\er})^{-1/2}$, and \er{} = \sr$/E = 0.0037$ (see more details in \SIi{}). 
	\fref{fig:material}e shows $Q_\text{beam}$ for beams of various lengths and in-plane orientations. 
	We obtain ${Q_\text{int} = (0.80\pm0.02)\times 10^4}$ for the \SI{290}{\nano \meter}-thick AlN film by fitting the data to \eref{eq:Qint}. 
	We use this value of \Qi{} as input for calculating the expected \Qd{} of various mechanical resonator geometries from finite element model (FEM) simulations.
	
	\subsection{Tapered phononic crystal beams}
\begin{figure*}[t!]
		\centering
		\includegraphics[width = \textwidth]{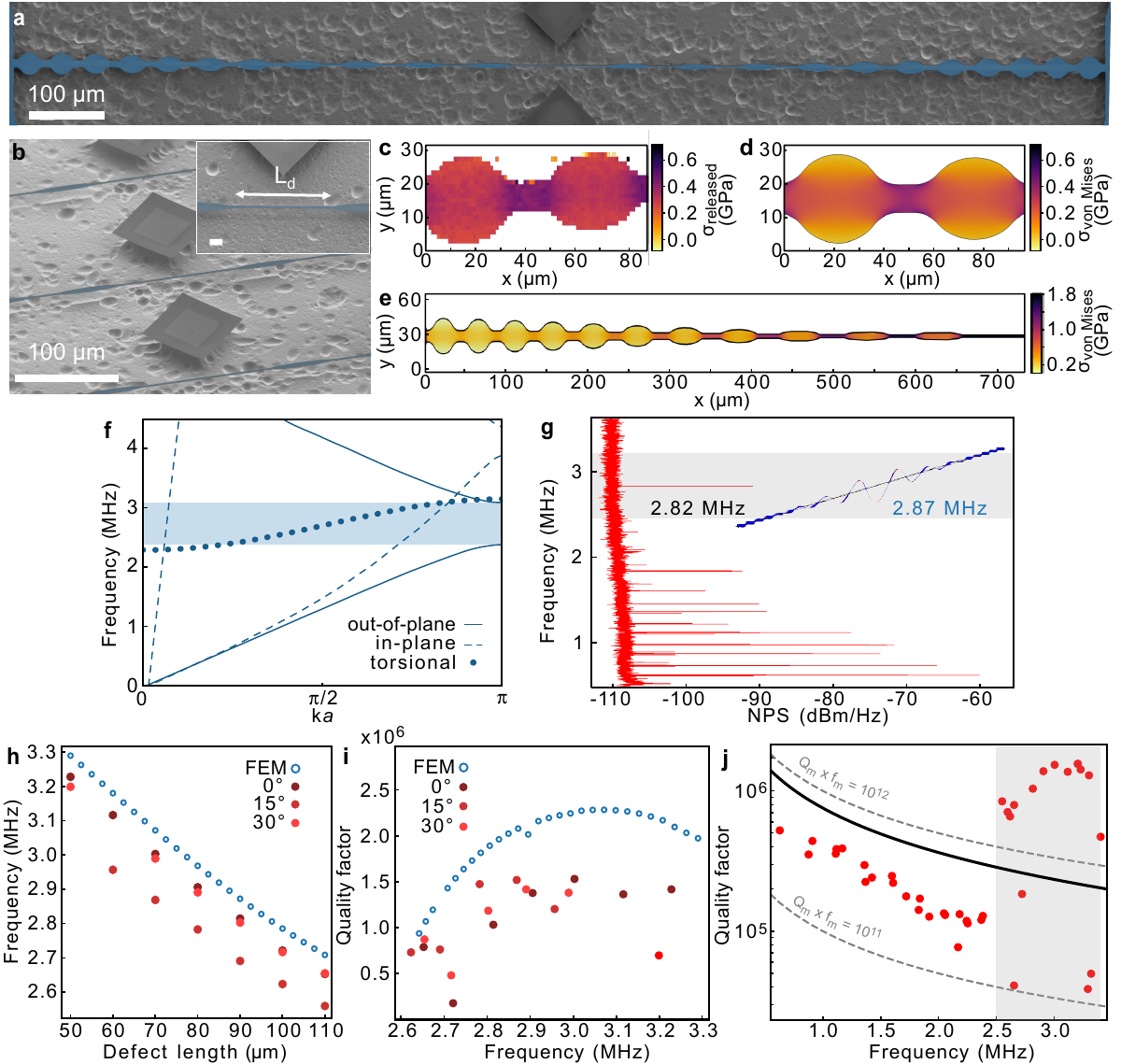}
		\caption{\textbf{Tapered phononic crystal beams in crystalline AlN.} a, b. False-colored SEM images of a \SI{290}{\nano \meter}-thick PnC beam. The inset in b shows the PnC defect of length $L_{d}$ = \SI{50}{\micro\meter} (scale bar \SI{10}{\micro \meter}). 
			c. Raman spectroscopy map of a part of the PnC unit cells $i = 9$ and $i = 10$. The colorbar shows \sr. d, e. FEM simulation results of the stress in the PnC beam for \sg\ = \SI{1}{\giga\pascal}. f. Band diagram of the phononic modes of the $i = 2$ unit cell for \sg\ = \SI{1}{\giga\pascal}.
			Modes are classified with respect to their transformation under the parity operation $(P_\textrm{y}, P_\textrm{z})$: $(1,-1)$  (solid lines), $(-1,1)$ (dashed lines), and $(-1,-1)$ (dotted lines), see \SIi{} for more details. g. Representative noise power spectrum (NPS). In the experiment we observe a defect mode at \SI{2.82}{\mega\hertz}, which is close to the FEM simulated value of \SI{2.87}{\mega\hertz}. h. Mode frequency in dependence of defect length and in-plane orientation of the PnC beam. i. Measured and simulated \Qm\ of the defect mode. j. Measured \Qm{} vs.~\fm{} of PnC beam modes, where defect modes show enhanced \Qm{} values, larger than uniform beams of similar frequency (solid black line).}
		\label{fig:1D_PnC}
	\end{figure*}

	Strained doubly-clamped beams exploit uniform stress for dissipation dilution, but they exhibit considerable bending at the clamping points leading to mechanical loss. Through the use of soft-clamped resonator designs \cite{tsaturyan2017ultracoherent,ghadimi2018elastic,fedorov2019generalized}, clamping-related bending losses can be eliminated. A straight-forward approach to implement soft clamping for a beam is to pattern it with a 1D PnC \cite{ghadimi2018elastic}. As a result, the \Qf-product of a defect mode in a PnC beam increases in comparison with uniform beams of similar frequency (see \fref{fig:1D_PnC}${\text{j}}$). 
	
	We pattern \SI{1.4}{\milli\meter}-long beams with a PnC. We additionally taper the width of the beam towards the center to co-localize a mechanical defect mode in the region of increased stress, which is a strain engineering method to further increase \Qm{} \cite{ghadimi2018elastic}. \fref{fig:1D_PnC}${\text{a, b}}$ show examples of fabricated devices.
	
	The unit cell of the 1D PnC has a length $a_\text{PnC}$ = 90\,$\mu$m at the tapered center and consists of a rectangular-shaped bridge of width $w_\textrm{min}$ = 2\,$\mu$m and an ellipse with a long axis of $a_\text{PnC}/3$ and short axis $a_\text{PnC}/27$. The sharp edges at their junction are rounded with a radius of 10\,$\mu$m. The rounded shapes are essential features as sharp corners that usually characterize these designs would lead to cracking of the highly stressed, brittle AlN film (see \SIi\ for more details). This unit cell is repeated 11 times in both directions from the center defect. We upscale the width of both the bridge (to realize tapering) and ellipse of the unit cells towards the clamping points. The width of the $i^{\mathrm{th}}$ iteration of the unit cell, scales as a Gaussian with $w(i)\sim \frac{1}{\beta} - \frac{(1-\beta)}{\beta} e^{- i^2/i_{0}^2}$ (with $-11\leq i \leq 11$ and $i_{0} = 7$, $\beta = 0.2$) \cite{ghadimi2018elastic}. At the same time we scale the length of the unit cell as $a_\text{PnC}(i) \sim 1/\sqrt{w(i)}$ to adapt the bandgap of each cell to the frequency of the defect mode \cite{ghadimi2018elastic}. 
	
	\fref{fig:1D_PnC}f shows the band diagram for the $i=2$ unit cell (\SIi{} shows band diagrams of additional unit cells). 
	We can classify the mechanical modes in this band diagram with regards to their parity under transformations $P_{l}$ where $l=x,y,z$, for example, $P_{x} (x^\prime,y^\prime,z^\prime)^\text{T} = (-x^\prime,y^\prime,z^\prime)^\text{T}$. 
	Such transformations applied to the displacement vector $\mathbf{u}$ yield, for example, $P_{x} \mathbf{u} (x,y,z)  = (P_{x} \mathbf{u})(P_{x} (x,y,z) ) = \pm \mathbf{u} (-x,y,z)$ \cite{Safavi-Naeini:10}. 
	We observe a bandgap between 2.5 and \SI{3}{\mega \hertz} for out-of-plane modes with $(P_\textrm{y},P_\textrm{z}) = (1,-1)$ symmetry. In-plane $(-1,1)$ and torsional $(-1,-1)$ modes cross this bandgap, but these modes do not couple to the out-of-plane motion due to symmetry (see \SIi{} for the simulated displacement of the modes and a description of their transformation under parity operations) \cite{tsaturyanUltracoherentSoftclampedMechanical2019}.
	
	The defect is formed at the middle of the PnC beam through the insertion of an additional bridge of length $L_{d}$ between the unit cells, see \fref{fig:1D_PnC}b, such that the defect mode frequency is within the effective bandgap \cite{ghadimi2018elastic}. This allows for soft-clamping of the defect mode and, at the same time, focuses the stress through the tapering of the bridges at the center. We perform Raman measurements of a fabricated PnC beam to assess its stress distribution. \fref{fig:1D_PnC}c shows results for two unit cells near the clamping point. 
	We observe a good match between the FEM simulated and measured \sr{} and \fm{} of the PnC beam for $\sigma_\text{residual,FEM}$ = \SI{1}{\giga\pascal}, \fref{fig:1D_PnC}c-d and \fref{fig:1D_PnC}h.
	The measured residual stress of the unreleased AlN film next to the PnC is, however, the same as on other samples, i.e., \sg{} = \SI{1.4}{\giga\pascal} (inferred from Raman measurements). The discrepancy between the measured residual stress, \sg{}, and the FEM assumed $\sigma_\text{residual,FEM}$ may be the result of inhomogeneous strain distribution of the AlN film, leading to partial stress relaxation and buckling of the ellipse regions in the PnC beam, thus, reducing locally its stress (see \SIi{}). 
	
	\fref{fig:1D_PnC}g shows a thermal noise displacement power spectrum (NPS) of a PnC beam with $L_{d}$ = 90\,$\mu$m. The defect mode is clearly visible at $f_{m} = \SI{2.82}{\mega\hertz}$, which is very close to the value obtained from FEM simulations. We summarize measurements of the defect mode frequency for PnC beams of different defect mode lengths and in-plane orientation in \fref{fig:1D_PnC}h (NPS of several beams are shown in \SIi).
	The frequency of the defect mode decreases with increasing defect length $L_{d}$, as expected from the FEM simulations, and is close to the simulated value. Furthermore, the defect mode frequency for the beam at \SI{0}{\degree} orientation and \SI{30}{\degree} are similar, while at \SI{15}{\degree} \fm{} is lower, see \fref{fig:1D_PnC}h. This observation is consistent with the in-plane angular dependence of \sr{} of the uniform beams (\fref{fig:material}d). 
	\fref{fig:1D_PnC}i shows \Qm{} of the defect modes. We observe no systematic change in the quality factor with beam orientation, similar to the in-plane rotated uniform beams (\fref{fig:material}e). While the trend of the measured \Qm{} versus defect mode frequency follows the trend of the FEM simulations, the absolute value of the measured data is slightly smaller than the FEM simulated one. This difference may be the result of buckling of the ellipses and fabrication imperfections (see \SIi), breaking the symmetry of localized defect modes and leading to mechanical dissipation through radiation loss into modes of other symmetry.
	
	\fref{fig:1D_PnC}j shows \Qm{} of delocalized and localized modes of the PnC beams. We observe that mechanical modes within the defect mode frequency range of \SI{2.5}{\mega \hertz} to \SI{3.5}{\mega \hertz} exhibit an enhanced \Qf-product. 
	We reach a maximal \Qf-product of up to $4.5\cdot 10^{12}$\,Hz, which is larger than the one of uniform beams of similar frequency (solid line in \fref{fig:1D_PnC}j). This confirms the soft clamping of the defect mode through the strain-engineered PnC.
	
	\subsection{Hierarchically-clamped triangline resonators}
	\begin{figure*}[htbp]
		\centering
		\includegraphics[width=\textwidth]{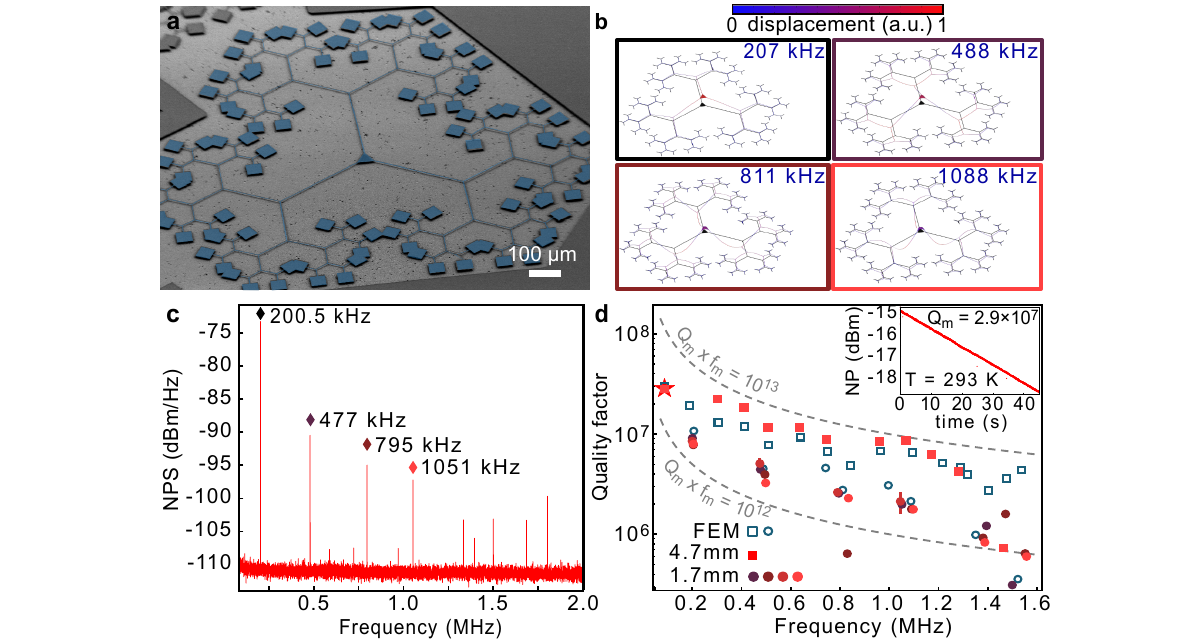}
		\caption{\textbf{Hierarchically-clamped triangline nanomechanical resonators in crystalline AlN.} a. False-colored SEM image of a triangline nanomechanical resonator with a total tether length of \SI{1.7}{\milli\meter}. b. FEM-simulated displacement and eigenfrequency of the first four eigenmodes of the triangline. c. Displacement NPS of the short triangline. The markers show the modes from panel b. d. Mechanical quality factor vs.~frequency for different trianglines: dots (squares) are short (long) trianglines. Filled markers are data, open markers are results from FEM simulations. The inset shows a ring-down measurement for the fundamental mode of a long triangline (red star) at \SI{87}{\kilo\hertz}.}
		\label{fig:triangline}
	\end{figure*}
	
	Optomechanical experiments with a Fabry-Pérot-type cavity require efficient and lossless coupling of a mechanical resonator's out-of-plane displacement to an optical beam. A mechanical resonator should then provide a non-absorbing, out-of-plane oscillating part with a sufficiently large area for accommodating an optical beam. 
	This can be realized, for example, with trampoline-like resonators \cite{norte2016mechanical,reinhardtUltralowNoiseSiNTrampoline2016,manjeshwar2023high}. To increase their \Qm{}, a type of soft clamping based on hierarchical structures can be applied \cite{fedorov2020fractal}. That has been demonstrated in Si$_3$N$_4$-based hierarchically-clamped trampolines \cite{bereyhi2022hierarchical} and trampoline-like geometries found by machine learning \cite{hoj2021ultra}. Further, the trampoline geometry typically realizes a high-\Qm{} for its fundamental mode, which is advantageous to use in certain quantum optomechanics protocols when nonlinear noise processes, such as thermal intermodulation noise \cite{fedorov2020thermal}, should be minimized. Finally, the central pad of a trampoline can be patterned with a PhC allowing to engineer its out-of-plane reflectivity \cite{fan2002analysis,norte2016mechanical,manjeshwar2023high}.
	
	A hierarchically-clamped trampoline consists of a central pad that is connected to four beams of length $l_{0}$. Each of these beams branches with an angle $\theta$ into two subsequent segments, with $N$ such branching iterations in total towards the clamping points \cite{fedorov2020fractal}. The length of subsequent segments ($1\leq n\leq N$) is $l_{n} = l_{0} r_{l}^{n}$ with $r_l<1$ and segment width $w = w_{0}(1/2\cos{\left( \theta \right)})^{n}$ to maintain a uniform stress in the structure \cite{fedorov2020fractal}.
	
	We first studied the effect of the branching angle on the \Qf-product of uniform-width beams in the crystalline AlN film. To this end, we fabricated beams with $N=1$ branching iteration and varied their branching angle $\theta$ (see \SIi). We find that $\theta = \SI{60}{\degree}$ yields singly-branched beams with a large \Qf-product. At the same time, a geometry with this branching angle follows the in-plane crystal structure of AlN, advantageous for achieving homogeneous stress along all directions of the branched tethers, resulting in uniform stress in all branched segments \cite{fedorov2020fractal}.
	
	To follow the in-plane $\SI{120}{\degree}$ rotation symmetry of the crystal, we use a triangular-shaped central pad that is suspended with three tethers. Each of the tethers branches off $N$ times at $\theta = \SI{60}{\degree}$ into two segments (see \fref{fig:triangline}a). We call this geometry triangline. When comparing this triangline with a hierarchically-clamped trampoline, we find that the expected \fm{} and \Qm{} of these two structures are very similar (see \SIi). An important advantage of the triangline geometry is that it enables a more economic use of the chip area, which allows us to increase the number of total branching iterations, and, thus, results in a larger \Qm{}.
	
	We fabricated two generations of trianglines that differ in their branching iterations $N$ and initial beam lengths $l_0$. Both generations have a central pad that is patterned with a PhC and has a side-length of 60\,$\mu$m. The pad is held by beams of constant width $w_{0}$ = 2\,$\mu$m and $r_{l} = 0.63$. We round the sharp edges of the triangline with a radius of 15\,$\mu$m to avoid cracking during fabrication.
	The first generation of trianglines, denoted as short, has $N=5$ branching iterations and an initial beam length of $l_{0}= \SI{0.32}{\milli \meter}$. This yields
	a total tether-length between furthest clamping points of \SI{1.7}{\milli\meter}.
	We simulate the first four eigenmodes and eigenfrequencies of this triangline, which are shown in \fref{fig:triangline}b. 
	The eigenfrequencies experimentally inferred from the measured thermal noise displacement power spectrum (\fref{fig:triangline}c) agree with FEM simulation results. 
	We measured \fm{} and \Qm{} of four devices and show the results in \fref{fig:triangline}d. 
	For the fundamental mode of the short triangline we find \fm{} $= \SI{200}{\kilo\hertz}$ with \Qm{} $= 9.4\times 10^6$, yielding a \Qf{}-product of $1.9\times 10^{12}$\,Hz. 
	We find that the FEM-predicted \fm{} and \Qm{} are in a good agreement with the measured values. Hence, we conclude that \Qm{} is limited by intrinsic dissipation (gas damping is not limiting dissipation mechanism, see measurements in \SIi).
	
	To increase \Qm{}, we fabricate a second generation of trianglines, which we denote as long, with one more branching iteration, i.e., $N=6$, and a longer initial beam length, $l_{0} = \SI{0.9}{\milli \meter}$, resulting in a total tether length of \SI{4.7}{\milli\meter} between furthest clamping points (see \SIi{}). This triangline exploits to the best of our knowledge the largest number of branching iterations demonstrated to date for trampoline-like hierarchically-clamped resonators \cite{bereyhi2022hierarchical}.
	The longer tether length lowers the fundamental mode eigenfrequency to \SI{87}{\kilo\hertz} while \Qm{} improves up to $2.9 \times 10^7$ (inset of \fref{fig:triangline}d), yielding a \Qf{}-product of $2.5\times 10^{12}$\,\SI{}{\hertz}, which is $30\%$ larger than for the fundamental mode of the short triangline. For higher order modes, the \Qf{}-product reaches $10^{13}$\,\SI{}{\hertz}, entering the regime of coherent oscillations for quantum optomechanics experiments at room temperature (required \Qf{}$>6\times 10^{12}$\,\SI{}{\hertz} \cite{aspelmeyer2014cavity}). We find a good agreement between FEM simulated and experimentally observed \fm{} and \Qm{} (\fref{fig:triangline}d).
	
	\begin{table*}[t!hbp]    
		\caption{\textbf{High-\Qm{} AlN nanomechanical resonators}. \fm{} and \Qm{} are experimental values for the mechanical mode eigenfrequency and mechanical quality factor, respectively. The motional mass, m$_\text{eff}$, is determined from FEM simulations, and the thermal force noise is calculated as $S_{F} = \sqrt{4k_B T \text{m}_\text{eff}\Gamma_m}$.}
		\label{tab:AlN_resonators}
		\begin{ruledtabular}
			\begin{tabular}{c|c c c c c c}
				& Length (\SI{}{\micro\meter}) &
				\fm (\SI{}{\kilo\hertz}) & \Qm & \Qf\  & 
				m$_\text{eff}$ (\SI{}{\nano\gram})  & 
				$\sqrt{S_F}$ (aN/$\sqrt{\text{Hz}}$) \\
				\hline
				\textbf{1D} & &  & & & & \\
				\hline
				uniform beam & 200 & 1400 & $5.5 \times 10^5$  & $0.77 \times 10^{12}$ & 0.047 & $111$\\
				Defect in 1D PnC & 80 & 3000  & $1.5 \times 10^6$  & $4.5\times 10^{12}$ & 0.35 & $269$\\
				\hline
				\textbf{2D} & &  & & & \\
				\hline
				PhC membrane & 180 & 2000 & $4.3 \times 10^5$  & $0.86\times 10^{12}$ & 0.9 & $657$\\
				Short triangline & 1700 & 200 & $9.4 \times 10^6$ & $1.88\times 10^{12}$ & 1.22 & $51.7$\\
				Long triangline & 4720 & 87 & $2.9 \times 10^7$  & $2.5\times 10^{12}$ & 2.67 & $29.2$\\
				\hline
				Long triangline & 4720 & 1068 & 
				$8.7 \times 10^6$  & 
				$9.2\times 10^{12}$ & 5.27 & 258.2\\
			\end{tabular}     
		\end{ruledtabular}    
	\end{table*}
	
	\begin{figure*}[t!hbp]
		\includegraphics[width = \textwidth]{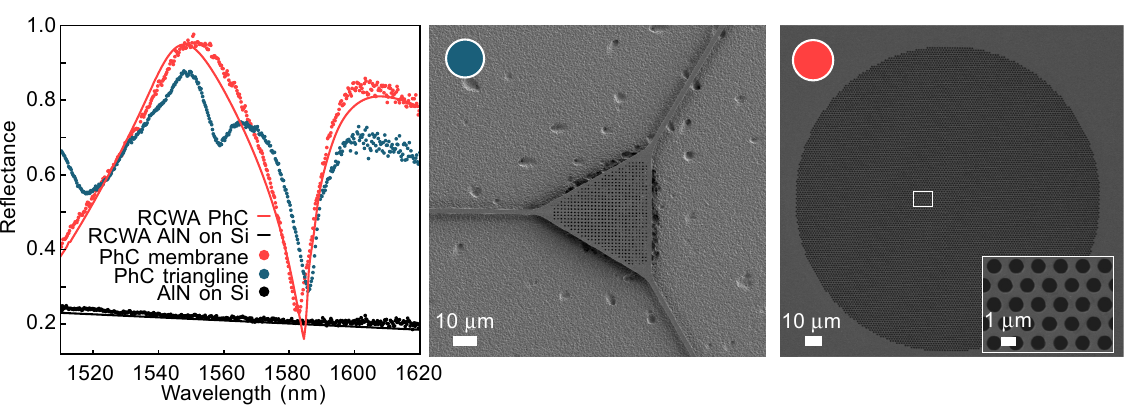}\
		\caption{\textbf{Engineering the reflectance of AlN nanomechanical resonators with a hexagonal PhC.}
			Reflectance vs.~wavelength measurements for an AlN film on a silicon substrate (black), a suspended \SI{180}{\micro\meter}-diameter circular membrane patterned with a PhC (red), and a triangline nanomechanical resonator with a center pad of \SI{60}{\micro\meter} side length patterned with a PhC (blue). Dots are experimental data, solid lines are RCWA simulations for a waist of \SI{6.4}{\micro\meter}. SEM images on the right: hexagonal PhC pattern in a suspended AlN nanomechanical resonator, scale bar \SI{1}{\micro\meter} (top right), triangline patterned with a PhC (middle right), circular membrane patterned with a PhC (lower right).}
		\label{fig:PhC}
	\end{figure*}
	
	The optical reflectivity of an unpatterned AlN film is determined by its thickness and refractive index. For the \SI{290}{\nano\meter}-thick AlN film at a wavelength of \SI{1550}{\nano\meter} the reflectivity is below 25\% (see \fref{fig:PhC}). To increase it, we pattern the central pad of the triangline with a PhC \cite{fan2002analysis,zhou2023cavity}, as shown in \fref{fig:PhC}.
	We pattern a hexagonal PhC with lattice constant $a_\text{PhC} = \SI{1450}{\nano\meter}$ and hole radius $r_\text{PhC} = \SI{508}{\nano\meter}$ into the AlN film to maximize its reflection of a normally-incident Gaussian beam at a wavelength of \SI{1550}{\nano\meter}. The PhC parameters were obtained from simulations of the suspended film's reflectivity using rigorous coupled-wave analysis (RCWA) \cite{liu2012s4} (see \SIi). 
	Note that the electro-optic effect in AlN could be used to in-situ tune the optical reflectivity with about a few picometers per volt, see \SIi{}.
	
	We patterned first a fully-clamped circular 180\,$\mu$m diameter AlN membrane with a PhC and observe that its reflectance is increased to above 90\%, see \fref{fig:PhC}. The circular clamping guarantees a uniform connection of the membrane to its support, but lowers \Qm{}, see \tref{tab:AlN_resonators}. 
	We observe pronounced reflectivity dips at \SI{1510}{\nano\meter} and \SI{1580}{\nano \meter}. 
	The first apparent dip originates from the Fano shape of the PhC reflectance.
	The second dip can be reproduced by RCWA simulations of the patterned \SI{290}{\nano\meter}-thick AlN film with an incident Gaussian beam of 6.4\,$\mu$m waist. 
	We thus identify the dip at \SI{1580}{\nano \meter} as coupling of the beam into a PhC guided resonance \cite{kini2020suspended}. 
	In the \SIi{} we show reflectance measurements with varying optical waist that let us clearly identify the guided resonance. Further, these measurements demonstrate that we can achieve a reflectance above 99\% when using a beam waist of 16\,$\mu$m. 
	We then patterned a PhC into the triangline's central pad and observe that its reflectance reaches a value above 80\% at \SI{1545}{\nano \meter}. Thus, the reflectance of the triangline's pad is drastically increased over the AlN film's reflectivity, but is slightly lower than the reflectance of the PhC-patterned circular membrane. 
	The radius of the PhC holes of the triangline and the circular membrane differs by \SI{6}{\nano\meter}. This causes a small shift of the overall Fano resonance to longer wavelengths. As a result, the PhC guided resonance appears now at \SI{1585}{\nano\meter} and the minimum of the Fano resonance at \SI{1515}{\nano\meter}.
	We also observe a reflectivity dip at \SI{1558}{\nano \meter}. 
	We identify this dip as the formation of an approximately 3.7\,$\mu$m-long low-finesse cavity between the PhC and the rough silicon substrate underneath. The overall lower reflectivity compared to the circular PhC membrane may be the result of the low-finesse cavity and the finiteness of the PhC lattice on the pad, which both influence the pad's reflectivity. 
	The reflectivity of the triangline's PhC pad could be improved by using a larger pad size to minimize finite-size effects.
	The triangline nanomechanical resonator could be integrated into a free-space optical cavity when incorporating some additional fabrication steps. For instance, one could introduce an AlInN/GaN-based distributed Bragg reflector and a sacrificial layer of GaN below the current AlN device layer during the growth. This would realize an optomechanical microcavity, similar to Ref.~\cite{KiniManjeshwar:23}. Alternatively, one could back-etch the Si substrate of the current devices to place the triangline nanomechanical resonator directly inside an optical cavity forming a membrane-in-the-middle-type optomechanical system \cite{gartner2018integrated}.
	
	\section{Conclusion and Outlook}\label{sec5}
	
	\tref{tab:AlN_resonators} summarizes the parameters of the soft-clamped nanomechanical resonators that we realized in \SI{290}{\nano\meter}-thick crystalline AlN. We calculated the thermal force noise to assess their performance in sensing applications. For example, with the fundamental mode of the long triangline we reach 29.2\,aN/$\sqrt{\text{Hz}}$, which is similar to conventional Si$_3$N$_4$-trampolines \cite{norte2016mechanical,reinhardtUltralowNoiseSiNTrampoline2016} or hierarchically-clamped Si$_3$N$_4$ trampolines \cite{bereyhi2022hierarchical}.
	
	\begin{table*}[tbh]
		\caption{
			\label{tab:comparison}
			\textbf{Materials used for high-\Qm{} nanomechanics at room temperature}. 
			The mechanical properties of the as-grown film are the residual stress \sg{}, Young's modulus $E$, and the strain $\epsilon$. The stress after relaxation, \sr, takes into account Poisson's ratio of the respective material.
		}
		\begin{ruledtabular}
			\begin{tabular}{l|cc|ccc|c}
				&
				Si$_3$N$_4$ \cite{bereyhi2019clamp} & SiC \cite{xu2023high}&
				Si\cite{beccari2022strained}& SiC \cite{romero2020engineering}&
				In$_{0.43}$Ga$_{0.57}$P\cite{manjeshwar2023high}
				&
				\textrm{AlN \footnote{This work.}}
				\\
				\colrule
				crystallinity & amorphous &amorphous & diamond cubic &  zinc blende & zinc blende & wurtzite \\
				thickness (nm) & 20 & 71 & 14 & 337 &73 & 290\\
				\Qi &$ 
				1.4 \times 10^3$
				& $5.1\times 10^3$ & $(8 \pm 3)\times 10^3$ & $10^4$ & $8\times 10^3$& $ 8 \times 10^3 $ \\
				\hline
				$n$ at \SI{1550}{\nano\meter} & 1.99 &  2.56 & 3.48  &  2.56 & 3.15 & 2.12\\
				\hline
				\sg (GPa)& 1.14& 0.76 & 1.53 $\pm$ 0.11 & 0.62 & 0.47 & 1.43$\pm0.01$\\
				$E$ (GPa)& 250 & 223 &169 & 400 & 80-120 & 270\\
				$\epsilon$ (\%) & 0.46 & 0.34 & 0.85 $\pm$ 0.06 & 0.15 & 0.49 & 0.36\\
				\sr (GPa)& 0.87 & 0.62 & 1 & 0.51 &0.3-0.5& 1\\
				\hline
				piezoelectricity, $\epsilon_\text{pol}$ \footnote{
					Wurtzite crystal: uniaxial strain along the $c$-axis $\epsilon_\text{pol} = \epsilon_{33}$.
					zinc blende: uniaxial strain along the $\langle 111 \rangle$ direction $\propto 2\epsilon_{14}/\sqrt{3}$ with an additional contribution from another polar direction \cite{hubner1973piezoelectricity}.        
				} (\SI{}{\coulomb\meter}$^{-2}$) \cite{hubner1973piezoelectricity,cimalla2007group} & - & - & - & $-1.24$& -0.23  
				& -1.55
				\\
				coupling coefficient, $k_\text{piezo}$ (\%) \cite{rais2014gallium} & - & - & - & 0.08 &  0.04 & 5.6\\
				relative permittivity, $\epsilon_r$ & 7-8 & 9.7 & 11.7 & 9.7 & 11.7 & 10\\
			\end{tabular}
		\end{ruledtabular}
	\end{table*}

	To put crystalline AlN into perspective with other materials used for high-\Qm{} nanomechanics, we summarize important material properties in Tab.~\ref{tab:comparison}. The residual stress of \SI{1.4}{\giga\pascal} of the AlN film is similar to state-of-the-art strained crystalline Si or amorphous Si$_3$N$_4$. 
	The AlN film's \Qi{} of $8 \times 10^3$ at room temperature is comparable to SiC or Si$_3$N$_4$ of similar thickness \cite{villanueva2014evidence}. 
	The \Qi{} of crystalline materials can, however, surpass the one of amorphous materials at cryogenic temperatures \cite{tao2014single,beccari2022strained}.
	The refractive index of AlN at telecom wavelengths is similar to that of Si$_3$N$_4$, but lower than for Si or InGaP.
	However, we demonstrated that the reflectance of the suspended AlN film can be vastly increased through patterning of a hexagonal PhC. Thanks to its bandgap of 6.2\,eV, and wide transparency window from the deep ultraviolet to mid infrared, AlN is an appealing material for low-loss quantum optical devices \cite{liu2023aluminum}. The use of AlN for quantum optomechanics devices operating at cryogenic temperatures may therefore improve device performance that is currently hampered by heating due to optical absorption. 
	
	A major advantage of using crystalline compared to amorphous materials for nanomechanical resonators is their in-built functionality. For example, crystalline films can be conducting or superconducting, or can exhibit piezoelectricity, provided they lack inversion symmetry. Zincblende or wurtzite crystals like SiC, InGaP and AlN meet the latter requirement. AlN has the largest piezoelectric coupling coefficient \cite{hubner1973piezoelectricity} among these materials (see \tref{tab:comparison}). 
	This has already been exploited in AlN-based GHz nanomechanics, where mechanical excitations, phonons, have been interfaced with superconducting qubits \cite{o2010quantum,mirhosseini2020superconducting}. 
	The piezoelectricity of the AlN film would allow, for example, to in-situ tune the mechanical frequency of the nanomechanical resonator. In our current devices we would expect a frequency tuning coefficient of some kHz/V, see \SIi{}. Parametric driving could then be used to generate squeezed mechanical states \cite{rugar1991mechanical}. To make use of the piezoelectricity of the AlN layer, a next step would be to alter the growth and microfabrication of the presented devices such that electrically conductive layers below and above the suspended AlN layer are incorporated.
	
	The presented triangline AlN nanomechanical resonators reached a \Qf{}-product close to $10^{13}$\,Hz (see \tref{tab:AlN_resonators}), sufficient to support a single quantum coherent oscillation at room temperature. We foresee multiple ways to increase device performance further by, for example, using other geometries, such as hexagonal polygon resonators \cite{bereyhi2022perimeter}, by etching the defect-rich layer \cite{romero2020engineering}, by reducing the overall thickness of the AlN film, or by operating at low temperatures. In particular, the thickness dependence of strain, intrinsic quality factor, and crystal quality are important parameters that will determine  the optimum working point for optoelectromechanical devices made from tensile-strained AlN \cite{ciers2024hvar}. In case of \SI{100}{\nano\meter}-thick AlN, one could potentially achieve \Qm{} of up to $10^{10}$ and a \Qf{}-product of $10^{15}$\,Hz (see \SIi), similar to crystalline Si nanomechanical resonators \cite{beccari2022strained}, but then realized in a piezoelectric material. Therefore, nanomechanical resonators from piezoelectric, tensile-strained AlN films hold great promise for interfacing kHz or MHz phonons with superconducting circuits for cavity quantum optomechanics \cite{brubaker2022optomechanical}, direct piezoelectric read-out of AlN-based nanomechanical resonators for sensing applications \cite{mahmoodi2022high}, or coupling of optical, mechanical and electronic degrees of freedom in the same material system \cite{midolo2018nano}.
	
	\clearpage
	\section{Methods}
	\subsection{Growth and fabrication details}
	The AlN film was grown with MOVPE (metal-organic vapour-phase epitaxy, AIXTRON AIX 200/4 RF-S) on a 2-inch 500\,$\mu$m-thick highly As-doped silicon (111) wafer.
	We used a two-step growth process, which is optimized to yield compact and very smooth AlN layers (typical AFM RMS $3 \times 3\, \mu$m$^2 < 0.2$ nm) free from pits \cite{dadgar2006movpe}. This is achieved by the second growth step which is performed under very low V/III ratio. However, applying such conditions right from the start of the growth of the layers turns out to result in a lower crystalline quality (as measured in XRD) and in a film that is more tensile-strained. To avoid this undesired feature, the initial 20 to 40\,nm of the AlN film is grown under a much higher V/III ratio.
	After a thin metallic Al deposition, \SI{20}{\nano\meter} of AlN were grown at a low growth rate with a surface temperature of \SI{1110}{\celsius}, \SI{100}{\milli\bar}, and a high V-III ratio of 2500. Then the main AlN layer was grown at \SI{70}{\milli\bar} and a low V-III ratio of 25 with a surface temperature of \SI{1120}{\celsius}.
	During the cooling process, strain is introduced due to the thermal expansion coefficient mismatch between the silicon substrate and the AlN film in addition to tensile strain during growth \cite{raghavan2004situ}. XRD measurements of the AlN film indicate a high crystalline quality (for details see \SIi{}).
	
	We start the fabrication process by sputtering a \SI{50}{\nano \meter} SiO$_2$ hard mask. Subsequently, we define the pattern of the mechanical resonator in electron-beam resist (UV-60). Then we transfer the pattern into the hard mask and AlN film in consequent ICP-RIE etching steps with CF$_4$/CHF$_3$ and Cl$_2$/Ar mixtures, respectively.
	We strip the photoresist with NMP (Remover 1165), and the sample is cleaned with one minute HF etching. To release the structure we use XeF$_2$ gas to selectively etch silicon. A mixture of XeF$_2$ and N$_2$ etches silicon isotropically with an etch rate of about 700\,nm/min \cite{winters1979etching}, whereas AlN is inert to XeF$_2$ at room temperature \cite{watanabe2005thermal}. During the release process the flux of hydrogen and nitrogen is set to 25\,sccm each and the pressure in the chamber is held at 1.2\,Torr.
	
	Note that the presented fabrication process was applied on the chip scale. However, both the growth as well as the fabrication process can be scaled up to four inch wafers at least.
	
	\subsection{Finite element simulations: mechanical properties}
	We use the solid mechanics interface of COMSOL Multiphysics for FEM simulations.
	First we find a stationary solution to determine the redistribution of \sg{}. The material parameters used for the simulations are listed in \tref{tab:param}.
	Then, we enter this static solution into the eigenfrequency solver. Finally, we extract the eigenfrequencies and the mechanical displacement field $u$, which we use further in the evaluation of \Qd{}.
	
	The dilution factor $D_{Q}$ is given by \cite{fedorov2019generalized}
	\begin{equation}
		D_{Q} = 1 + \frac{\left< \Delta W^\textrm{(nl)} \right>}{\left< \Delta W^\textrm{(lin)} \right>},
		\label{eq:dissipation_dilution_detail}
	\end{equation}
	where the time-averaged $\left< \Delta W^\textrm{(lin)} \right>$ and $\left< \Delta W^\textrm{(nl)} \right>$ are the linear and non-linear dynamic contributions to the elastic energy of a specific mode shape, respectively.
	
	To calculate $D_Q$ of a crystalline mechanical resonator of thickness $h$ we resort to the general relation between the components of the stress tensor $\sigma_{ij}$ and the strain tensor $\epsilon_{kl}$ through the elasticity matrix $C_{ijlk}$, i.e., $\sigma_{ij} = C_{ijlk} \epsilon_{kl}$. We use this relation and calculate an expression for the time-averaged linear elastic energy $\left< \Delta W^\textrm{(lin)} \right>$,
	
	\begin{align}
		\left< \Delta W^\textrm{(lin)} \right> =& \frac{h^{2}} {24} \iint\limits_V \Biggl(
		\left(
		\frac{\partial^2{u}}{\partial x^2} \right) ^2
		\left[C_{11}-\frac{C_{13}^2}{C_{33}} \right] +
		\nonumber\\ 
		& + 2 \frac{\partial^2{u}}{\partial x^2}\frac{\partial^2{u}}{\partial y^2}   \left[C_{12}-\frac{C_{13}C_{23}}{C_{33}}\right] + \nonumber\\
		& + \left(
		\frac{\partial^2{u}}{\partial y^2} \right) ^2 \left[C_{22}-\frac{C_{23}^2}{C_{33}}\right] +
		\label{eq:density_bending_energy_anisotropic}\\ 
		& + 4 C_{66}
		\left( \frac{\partial^2{u}}{\partial x\partial y} \right) ^2 \Biggr) dV, \nonumber
	\end{align}
	
	where $u$ is the out-of-plane displacement and the integral $V$ covers the volume of the mechanical resonator geometry. For details with regards to this calculation we refer to the \SIi.\footnote{We supply a short Mathematica script for these calculations on linear contribution to the elastic energy for an anisotropic material \cite{zenododata}.}
	In particular, this treatment covers isotropic materials and crystals with cubic, hexagonal and orthorhombic crystal systems.
	We use \eref{eq:density_bending_energy_anisotropic} to calculate $\left< \Delta W^\textrm{(lin)} \right>$ for a given eigenmode of the resonator obtained from the FEM simulations.
	The expression for the calculation of the time-averaged non-linear contribution is related to the total energy of the system and remains unchanged
	\begin{equation}
		\left< \Delta W^\text{(nl)} \right> = \rho \frac{ \omega ^2}{2} \iint\limits_V u^2 dV,
		\label{eq:nonlinear_contribution} 
	\end{equation}
	where $\omega$ is the eigenfrequency of the particular eigenmode \cite{fedorovMechanicalResonatorsHigh2021}. We also evaluate this expression for a given eigenmode in the FEM simulations.
	We can subsequently calculate the dilution factor $D_{Q}$ with \eref{eq:dissipation_dilution_detail} and obtain $Q_\text{D}=D_{Q}\cdot Q_\text{int}$.
	
	We determine the motional mass as
	\begin{equation}
		m_\textrm{eff} = \rho \frac{\int_{V} \left| u \right| ^2 dV}{ \left| u_\textrm{max} \right| ^2},
		\label{eq:motional_mass}
	\end{equation}
	where we integrate over the entire geometry with the displacement $u$ and the maximum displacement $u_\textrm{max}$ of that particular eigenmode.
	
	\begin{table}[h]
		\begin{center}
			\caption{Parameters for the FEM simulation. AlN film thickness, $h$, density $\rho$, residual stress, \sg, intrinsic quality factor, \Qi, elastic constants, $C_{ij}$ \cite{tsubouchi1981aln}. We calculate the effective Young's modulus, $E$, and Poisson's ratio, $\nu$, with $C_{ij}$.}
			\label{tab:param}
			\begin{tabular*}{250pt}{@{\extracolsep\fill}lc@{\extracolsep\fill}}%
				\toprule
				thickness, $h$ (nm) & 290 \\
				density, $\rho$  (kg/m$^3$) & 3255 \\ 
				\sg{} (GPa) & 1.4\footnote{Note, that \sg{} differs for the PnC beam simulations and is \SI{1}{\giga \pascal}.}\\
				\Qi{} & $0.8\times 10^4$\\
				\hline
				$C_{11}$ (GPa) & 345 \\
				$C_{33}$ (GPa) & 395\\
				$C_{44}$ (GPa) & 118\\
				$C_{12}$ (GPa) & 125\\
				$C_{13}$ (GPa) & 120\\
				\hline
				$E$ (GPa) & 283\\
				$\nu$ & 0.287\\
				\hline
				\hline
			\end{tabular*}
		\end{center}
	\end{table}
	
	The values of the elastic constants are chosen from the measurements on AlN films \cite{tsubouchi1981aln} with the resulting Young's modulus close to the one determined in this work, \SI{270}{\giga\pascal}. It should be noted that the difference between the reference parameters \cite{tsubouchi1981aln} and the presented AlN film could be a reason for the small discrepancies of \Qd{} between the FEM simulation and the experiment.
	
	\subsection{Optical reflectance simulation}
	We use a rigorous coupled wave analysis (RCWA) solver, which is available as the Stanford Stratified Structure Solver ($S^4$) software package \cite{liu2012s4}. The focused Gaussian beam with a waist is reconstructed as a superposition of plane waves impinging on the PhC \cite{kini2020suspended}. The value of the PhC parameters ($a_\text{PhC}$, $r_\text{PhC}$) is determined after fabrication via analysis of SEM images. The radius of the PhC holes in the PhC pattern of a circular membrane is $r_\text{PhC} = \SI{508}{\nano\meter}$ and in the case of the triangline it is $r_\text{PhC} = \SI{502}{\nano\meter}$. The small difference in $r_\text{PhC}$ results in the slightly different position of the guided mode at \SI{1585}{\nano\meter} observed in \fref{fig:PhC}.
	
	\subsection{Raman spectroscopy}
	\label{subsec:Raman}
	Residual stress and strain of the AlN film is evaluated by means of Raman scattering using a confocal Raman microscope at room temperature.
	A \SI{532}{\nano\meter} laser is used for excitation and was focused down by an objective lens with a magnification of 100$\times$ and a numerical aperture of 0.9, leading to an optical spot size of about \SI{300}{\nano \meter}. We verified that the Raman signal is power independent and, thus, not affected by a potential heating of the device. Note that the calculation that relates the wavelength of the Raman line to the stress and strain in the AlN film is outlined in the \SIi{}.
	
	\subsection{Interferometric characterization setup}
	\label{subsec:Characterization}
\begin{figure}[h]\
		\centering
		\includegraphics{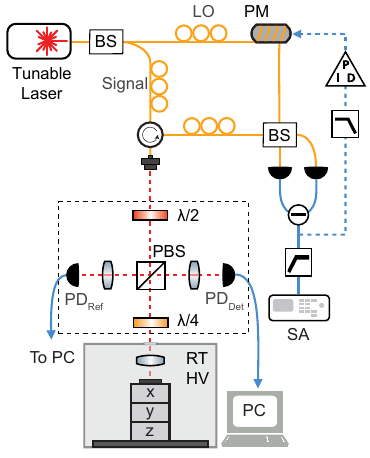}\
		\caption{The experimental setup consists of a homodyne detection and a reflectivity measurement (in black-dashed box) parts. The fiber path (yellow solid), free-space (dashed red) and an electrical connections (blue).}\
		\label{fig:setup}
	\end{figure}

	The measurements of AlN resonators are performed at $7\times10^{-6}$\,mbar and at room temperature, using optical interferometry driven by a tunable laser with a wavelength of 1550\,nm, see \fref{fig:setup}.
	\
	
	The laser beam is reflected off the sample inside a vacuum chamber, and we detect the resulting phase shift using a phase-locked homodyne detector and record its output signal with a spectrum analyzer (SA). The ring-down measurements are performed by resonantly driving the sample with a piezoelectric transducer, switching it off and then recording the decay of the signal.
	
	To measure the reflectance of the samples, we adjust the polarization with a half-wave plate before the polarizing beam-splitter (PBS) and monitor the incoming laser power via a reference photodetector, $\text{PD}_\text{Ref}$. The output from $\text{PD}_\text{Ref}$ is fed back to the laser to stabilize the laser power throughout the wavelength-sweep.
	After the PBS we insert a quarter-wave plate to detect the reflected intensity with the $\text{PD}_\text{Det}$ photodetector.
	
	\begin{acknowledgments}
		We thank Nils Johan Engelsen for valuable discussions and comments on the manuscript and Alexei Kalaboukhov for support with piezoresponse force microscopy measurements. This work was supported by the Knut and Alice Wallenberg (KAW) Foundation through a Wallenberg Academy Fellowship (W.W.), the KAW project no.~2022.0090, and the Wallenberg Center for Quantum Technology (WACQT, A.C.), the Swedish Research Council (VR projects No. 2019-04946), the QuantERA project C’MON-QSENS!, and Chalmers’ Area of Advance Nano. H.P. acknowledges funding by the European Union under the project MSCA-PF-2022-OCOMM. MOVPE of AlN on Si was performed at Otto-von-Guericke-Univerrsity Magdeburg. The mechanical resonators were fabricated in the Myfab Nanofabrication Laboratory at Chalmers and analyzed in the Chalmers Materials Analysis Laboratory. Simulations were performed on resources provided by the National Academic Infrastructure for Supercomputing in Sweden (NAISS) partially funded by the Swedish Research Council through grant agreement no. 2022-06725.
	\end{acknowledgments}
	
	\section*{Data availability}

	Data underlying the results presented in this paper are available in the open-access Zenodo database: https://doi.org/10.5281/zenodo.10679216  \cite{zenododata}.
		
	\clearpage
	
	%
	
	\appendix
	\section*{Supplementary Material}
	
	\setcounter{figure}{0}
	\setcounter{equation}{0}
	\setcounter{section}{0}
	\renewcommand{\thefigure}{S\arabic{figure}}
	\renewcommand{\theequation}{S\arabic{equation}}
	
	\section{Material characterization}
	\subsection{X-ray diffraction}
	\label{sec:xrd}
	X-ray diffraction (XRD) is a technique used to analyze the crystallographic structure of materials by observing how X-rays scatter off the crystal lattice. This data allows determining the crystal structure, crystalline quality, identify phases, calculate lattice parameters, and assess the size and strain of crystalline domains.
	The full-width-at-half-maximum (FWHM) of the (0002) reflex in XRD rocking curves quantifies the crystallinity of the AlN film. A smaller FWHM of this reflex corresponds to a more uniform alignment of AlN crystal domains along the $c$-axis.
	
	\begin{figure*}[h!tbp]
		\centering
		\includegraphics[width = \textwidth]{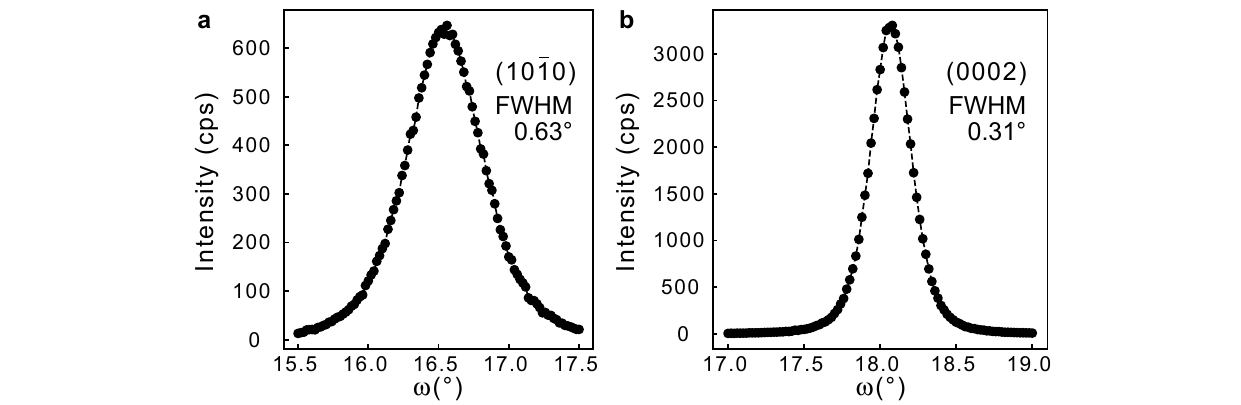}
		\caption{XRD data a. (10$\bar{1}$0) peak, b. (0002) peak with corresponding FWHM.}\
		\label{fig:XRD}
	\end{figure*}
	\fref{fig:XRD} shows the (10$\bar{1}$0) and (0002) XRD peaks of the MOVPE-grown AlN film on Si (111). We determine a FWHM of the (0002) reflex of \SI{0.31}{\degree}. This value is, for example, smaller than the one observed for a sputtered polycrystalline AlN film that exhibits a FWHM above \SI{2}{\degree} \cite{howell2019effect}, indicating a better crystallinity of our film.
	
	\subsection{Transmission electron microscopy}
\begin{figure}[h]
		\includegraphics[width = 0.48\textwidth]{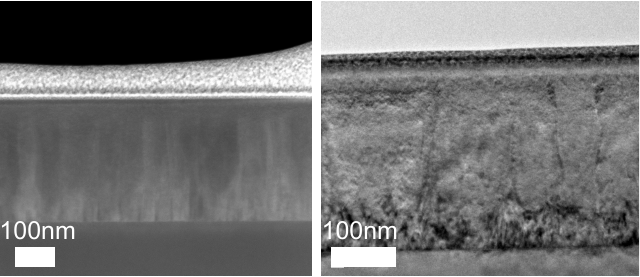}\
		\caption{TEM and STEM images of the \SI{290}{\nano\meter}-thick AlN film on Si substrate.}\
		\label{fig:TEM}
	\end{figure}

	We perform TEM imaging to verify the thickness and quality of the MOVPE-grown AlN film on Si (111). 
	A large dislocation density is found in the interface layer between AlN and the Si substrate. Further away from the substrate, the AlN forms regions of dislocations and misoriented domains. However, the quality of AlN improves drastically with increasing AlN film thickness, as seen in the TEM image in \fref{fig:TEM}.
	
	\subsection{Ellipsometry}
	\begin{center}
		\begin{figure}[h]
			\includegraphics[width=0.48\textwidth]{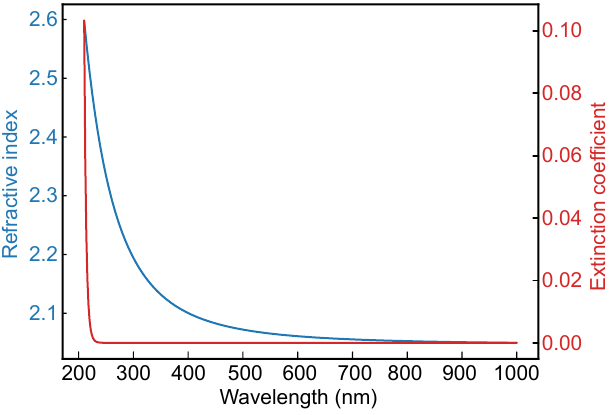}
			\caption{Refractive index and extinction coefficient of the \SI{290}{\nano\meter}-thick AlN film.}
			\label{SI:fig_ellipse}
		\end{figure}
	\end{center}
	
	We measured the refractive index and extinction coefficient of the AlN film via ellipsometry, Woollam RC2. The fitted results are in \fref{SI:fig_ellipse}. The obtained values are in a good agreement with the literature \cite{pastrvnak1966refraction}.

	\subsection{Atomic force microscopy}
	We performed AFM measurements with an SPM Bruker Dimension 3100 in tapping mode at \SI{7}{\micro\meter/\second} speed. The surface roughness of the AlN film prior to fabrication is $R_q$(RMS)$ = 1 \pm 0.1$\,nm, which is comparable to the previously reported values of similarly grown MOCVD AlN films on Si(111) \cite{dai2016properties}. In \fref{fig:AFM} one can see a representative \SI{5}{\micro\meter}$\times$\SI{5}{\micro\meter} scan.  
	
	\begin{center}\
		\begin{figure}[h]
			\includegraphics[width=0.4\textwidth]{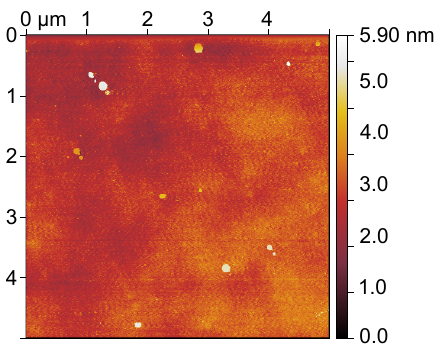}\
			\caption{Surface topography of AlN film scanned with AFM: $R_q$(RMS) = \SI{950.5}{\pico\meter}}\
			\label{fig:AFM}
		\end{figure}\
	\end{center}\ 
	
	\subsection{Elastic anisotropy}
	The elastic anisotropy of materials has a significant effect on their physical properties, such as deformation and crack propagation.
	In the following we provide equations in the case of plane stress for isotropic, cubic, hexagonal and orthorhombic cases. 
	This is done to convert the results of Raman spectroscopy to the residual stress of the film and to derive the dissipation dilution factor. Then we show how to calculate uniaxial stress, which corresponds to measured stress of a suspended beam in Raman spectroscopy.
	
	In the following the $z$-axis coincides with the $c$-axis of AlN and the $x$, $y$ axes lie in the $m_1m_2$-plane.
	
	\subsubsection{General case}
	First, we take a general elasticity matrix which describes isotropic, cubic, hexagonal and orthorhombic cases:
	\begin{equation}
		\begin{pmatrix}
			\sigma_{xx}\\
			\sigma_{yy}\\
			\sigma_{zz}\\
			\sigma_{yz}\\
			\sigma_{zx}\\
			\sigma_{xy}\\
		\end{pmatrix}
		= 
		\begin{pmatrix}
			C_{11} & C_{12} & C_{13} & 0  & 0 & 0  \\
			C_{12} & C_{22} & C_{23} & 0  & 0 & 0  \\
			C_{13} & C_{23} & C_{33} &  0 & 0 & 0  \\ 
			0 & 0 & 0 & C_{44} & 0 & 0 \\
			0 & 0 & 0 & 0 & C_{55} & 0 \\
			0 & 0 & 0 & 0 & 0 & C_{66}
		\end{pmatrix}%
		\begin{pmatrix}
			\epsilon_{xx}\\
			\epsilon_{yy}\\
			\epsilon_{zz}\\
			2\epsilon_{yz}\\
			2\epsilon_{zx}\\
			2\epsilon_{xy}\\
		\end{pmatrix},
		\label{eq:C_general}
	\end{equation}
	Then, in the case of a thin membrane of thickness $h$, we obtain a plain stress condition, where the $z$-components of the tensor are zero:
	\begin{equation}
		\sigma_{xz} = \sigma_{yz} = \sigma_{zz} = 0.
		\label{eq:stress_plane}
	\end{equation}
	Applying \eref{eq:stress_plane} to the stiffness matrix \ref{eq:C_general} leads to the following strain relations:
	\begin{equation}
		\epsilon_{zz} = -\frac{C_{13}\epsilon_{xx}+C_{23}\epsilon_{yy}}{C_{33}},\ \epsilon_{xz}=\epsilon_{yz} = 0.
		\label{eq:condition_anisotropic}
	\end{equation}
	
	\subsubsection{Hexagonal symmetry}
	The elasticity matrix of the hexagonal crystal is described by 5 elastic constants (see Table \MakeUppercase{\romannumeral 3}) \cite{mouhat2014necessary}:
	\begin{equation}
		\begin{pmatrix}
			\sigma_{xx}\\
			\sigma_{yy}\\
			\sigma_{zz}\\
			\sigma_{yz}\\
			\sigma_{zx}\\
			\sigma_{xy}\\
		\end{pmatrix}
		= 
		\begin{pmatrix}
			C_{11} & C_{12} & C_{13} & 0 & 0 & 0\\
			C_{12} & C_{11} & C_{13}& 0 & 0 & 0\\
			C_{13} & C_{13} & C_{33}& 0 & 0 & 0\\
			0 & 0 & 0 & C_{44} & 0 & 0\\
			0 & 0 & 0 & 0 & C_{44} & 0\\
			0 & 0 & 0 & 0 & 0 & C_{66}\\
		\end{pmatrix}%
		\begin{pmatrix}
			\epsilon_{xx}\\
			\epsilon_{yy}\\
			\epsilon_{zz}\\
			2\epsilon_{yz}\\
			2\epsilon_{zx}\\
			2\epsilon_{xy}\\
		\end{pmatrix},
		\label{mat:hex}
	\end{equation}
	and the Cauchy relation
	\begin{equation}
		C_{66} = \frac{ C_{11} -  C_{12}}{2}.
	\end{equation}
	The compliance matrix $S$ is the inverse of the elasticity matrix:
	\begin{widetext}
	\begin{equation}
		S = C^{-1} = 
		\frac{1}{\Delta}
		\begin{pmatrix}
			C_{11}C_{33}- C_{13}^2 & C_{13}^2 - C_{12}C_{33} & (C_{12}-C_{11})C_{13} & 0 & 0 & 0\\
			C_{13}^2 - C_{12}C_{33} & C_{11}C_{33}- C_{13}^2 & (C_{12}-C_{11})C_{13} & 0 & 0 & 0\\
			(C_{12}-C_{11})C_{13} & (C_{12}-C_{11})C_{13} & C_{11}^2 - C_{12}^2& 0 & 0 & 0\\
			0 & 0 & 0 & 1/C_{44} & 0 & 0\\
			0 & 0 & 0 & 0 & 1/C_{44} & 0\\
			0 & 0 & 0 & 0 & 0 & 1/C_{66}\\
		\end{pmatrix}%
		\label{mat:S_hex}
	\end{equation}
	\end{widetext}
	where $\Delta = (C_{11} - C_{12})\left((C_{11} + C_{12})C_{33} - 2C_{13}^2\right)$.
	
	For the hexagonal crystal, Young's modulus and Poisson's ratio are isotropic in the plane perpendicular to the $c$-axis.
	
	\subsubsection{Biaxial stress}
	The as-grown hexagonal membrane is under equi-biaxial stress, $\sigma_{xx} = \sigma_{yy}$.
	Then the stress tensor can be derived from the matrix \eref{mat:hex} as
	\begin{equation}
		\sigma_{xx} = \sigma_{yy} = (C_{11} + C_{12})\epsilon_{xx} + C_{13}\epsilon_{zz}
	\end{equation}
	\begin{equation}
		\sigma_{zz} = 2 C_{13}\epsilon_{xx} + C_{33}\epsilon_{zz}.
	\end{equation}
	
	For a thin membrane we apply the plane stress condition\footnote{$\sigma_{ij}^0, \epsilon_{ij}^0$ denote the stress and strain, respectively, for the case of plane stress in a membrane out of material with hexagonal symmetry}, \eref{eq:stress_plane} ($\sigma_{zz}^0 = 0$). Then one obtains the stress- and strain-component relationship for a hexagonal crystal as
	\begin{equation}
		\epsilon_{zz}^0 = -2\frac{C_{13}}{C_{33}}\epsilon_{xx}^0
		\label{eq:Raman_epsilon},
	\end{equation}
	\begin{equation}
		\sigma_{xx}^0 = \left(C_{11} + C_{12} - 2\frac{C_{13}^2}{C_{33}}\right)\epsilon_{xx}^0,
		\label{eq:Raman_sigma}
	\end{equation}
	which we use further to evaluate the residual stress of the AlN film on the Si substrate from Raman measurements.
	
	\subsubsection{Uniaxial stress}
	For uniaxial stress\footnote{$\sigma_{ij}^1, \epsilon_{ij}^1$ denotes the stress and strain, respectively, for the case of $\sigma_{zz}^1 = 0$ in a beam}, $\sigma_{xx}^1$, as in the case of the beam, $\sigma_{yy}^1$ vanishes and, hence, an elastic relaxation occurs. The resulting in-plane strain is expressed by the Poisson ratio and is defined by the elastic constants in Tab.\MakeUppercase{\romannumeral 3}.
	The strain components are calculated as
	\begin{equation}
		\epsilon_{zz}^1 = \sigma_{xx}^1\frac{C_{13}}{C_{33}(C_{11} - C_{12})},
		\label{eq:uniax_strain_z}
	\end{equation}
	\begin{equation}
		\epsilon_{xx}^1 = \sigma_{xx}^1\frac{C_{11}C_{33} - C_{13}^2}{(C_{11} - C_{12})(C_{33}(C_{11} + C_{12}) - 2C_{13}^2)},
		\label{eq:uniax_strain_x}
	\end{equation}
	\begin{equation}
		\epsilon_{yy}^1 = \epsilon_{zz}^1\frac{C_{13}^2 - C_{12}C_{33}}{C_{13}(C_{11} - C_{12})}.
		\label{eq:uniax_strain_y}
	\end{equation}
	
	We observed in the experiment, \sr{}$=\sigma_{xx}^1$ is anisotropic, which would require us to lower the symmetry from hexagonal to trigonal and introduce $C_{14}$ to obtain \SI{60}{\degree}-periodicity. 
	This is expected, as wurtzite AlN is a hexagonal system, however, the 6-fold axis is absent. 
	This means that if we rotate the AlN crystal around $c$-axis by \SI{60}{\degree} (top view Fig.~1b in the main text) aluminum atoms will be above nitrogen position and therefore the AlN crystal is not reproduced upon this transfomation.
	As predicted by Neumann’s Principle states \cite{hartmann1984introduction}: “The symmetry of any physical property of a crystal must include the symmetry elements of the point group of the crystal", hence one would anticipate the weak $c$-plane anisotropy of \sr{}.
	An important work \cite{zeng2017crystal} on GaN showed a strong crystallographic orientation dependence of the sliding properties with a \SI{60}{\degree} periodicity of wear rate and friction coefficient.
	The anisotropy in friction comes from the energetic barriers derived from the crystalline structure that governs wear.
	Such wear tests are, in essence, the measure of the material change with application of a uniaxial force. 
	Therefore, if the beam relaxes anisotropically this leads to anisotropic \sr{}.
	
	Nevertheless, in the stress, strain and FEM evaluation we assume hexagonal symmetry of AlN, as the anisotropy is weak and only 5 elastic constants are available in literature.
	Using the experimental value for \sr{} = \SI{1}{\giga\pascal} and Eq.~\ref{eq:uniax_strain_x}, we find that the released strain of the beam is $\epsilon_{xx}^1 \approx 0.0035$. This value is close to \sr{}$/E$ = 0.0037 defined from the beam frequencies and we use this value in the calculation of \Qi{}.
	
	\subsection{Raman spectroscopy}
	\begin{figure}[h!tbp]
		\includegraphics[width=0.45\textwidth]{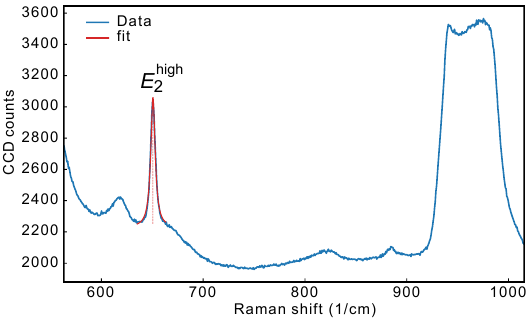}
		\caption{Raman shift of \SI{290}{\nano\meter} AlN film with fit of $E_2^\text{high}$ at 650.67$\pm 0.03$
			\,cm$^{-1}$.}
		\label{fig:Raman_290}
	\end{figure}
	
	In the experiment we have access to the strained unreleased AlN film on Si (equi-biaxial case, \eref{eq:Raman_epsilon}) and released AlN beams (uniaxial case). Prior measurements we verify that the Raman spectrum doesn't depend on the laser power.
	We start by measuring equi-biaxial strain and then link it to uniaxial strain in the beam.
	
	In \fref{fig:Raman_290}, one can see that the AlN film exhibits $E_2^\text{high}$ peak at 650.67\,cm$^{-1}$ and $A_1$ (TO) at 618\,cm$^{-1}$.
	
	As the chemical bond length increases, while the force constant remains the same, the vibrational frequency decreases. In tensile-strained materials the Raman peak position is shifted to lower frequencies \cite{dai2016properties}. As the relaxed value of the $E_2^\text{high}$ AlN phonon frequency we take the bulk AlN crystal value of 657\,cm$^{-1}$ \cite{zhuang2004bulk}.
	
	In the equi-biaxially strained case, a frequency shift $\Delta\omega$ is determined by the deformation potential constant $\alpha_0$ as \cite{callsen2014phonon}
	\begin{equation}
		\Delta\omega = 2\alpha_0 \sigma_{xx}^0
	\end{equation}
	where $2\alpha_0 = 4.423$\,cm$^{-1}$/GPa \cite{dai2016properties} for AlN and one can directly evaluate \sg{} $= 1.43\pm \SI{0.01}{\giga\pascal}$. Then it is possible to determine the in-plain strain using \eref{eq:Raman_sigma} and \eref{eq:Raman_epsilon} to be ${\epsilon_{xx}^0 = 0.0036}$ and ${\epsilon_{zz}^0 = 0.0021}$. 
	The released stress of the beam is related to the residual stress as ${(1 - \nu) \sigma_\text{residual} = \SI{1.02}{\giga\pascal}}$, which matches well with the fitted \sr{} obtained from measurements of beams presented in the main text. 
	
	By mapping the fitted value of $E_2^\text{high}$ on the suspended beams we observe the local strain variation in suspended AlN. 
	\begin{figure}[h]
		\includegraphics[width=0.45\textwidth]{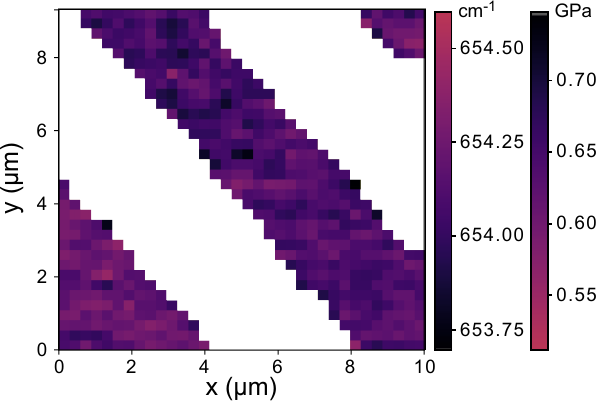}\
		\caption{Raman spectroscopy map of \SI{290}{\nano\meter} AlN uniform \SI{2}{\micro\meter}-wide beam. The left colorbar is $E_2^\text{high}$ frequency, the right colorbar is \sr.}
		\label{fig:Raman_uni}
	\end{figure}
	In case of the uniform beam the released $E_2^\text{high}$ mode position remains constant, see \fref{fig:Raman_uni}. 
	While with the width variation in the PnC beam, the strain is higher at the narrow parts and approaches the suspended uniform beam $E_2^\text{high}$ frequency, as can be seen in Fig.~2 in the main text. At the wide parts of the PnC beam, i.e., the elliptic wings, AlN is relaxed and the $E_2^\text{high}$ mode is at the bulk crystal value of 657\,cm$^{-1}$.
	
	\subsection{Young's modulus determination}
	We used \SI{200}{\micro \meter}-long uniform beams to evaluate the Young's modulus of the AlN film following the method presented in Ref.~\cite{klass2022determining}. We measured higher-order mode frequencies ($n,m$) of the beams. Young's modulus can then be obtained as
	\begin{equation}
		E = \frac{48L^4\rho}{\pi^2h^2(n^2-m^2)}\left( \frac{f_n^2}{n^2} - \frac{f_m^2}{m^2}\right),
	\end{equation}
	where $f_{n,m} $ are the higher-order mode frequencies \cite{klass2022determining}.
	
	As crystalline materials demonstrate elastic anisotropy \cite{manjeshwar2023high, hopcroft2010young}, we performed measurements for beam in-plane orientations $\alpha = 0^\circ$, \SI{60}{\degree}, \SI{90}{\degree} and \SI{120}{\degree}.
	
	\begin{figure}[h]
		\includegraphics[width = 0.47\textwidth]{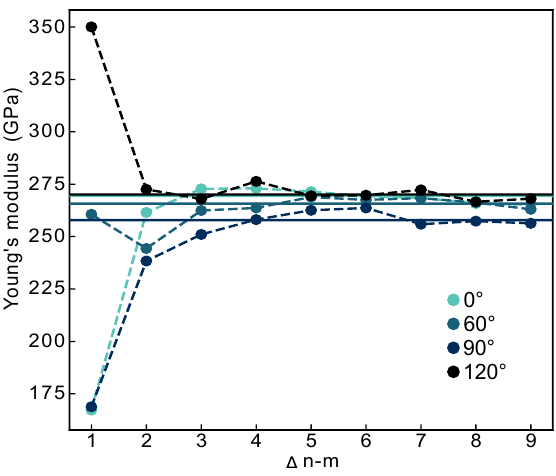}\
		\caption{Young's modulus of a 290\,nm-thick crystalline AlN film. The solid lines mark the mean value that is evaluated from all mode differences above 2.}
		\label{fig:E}
	\end{figure}
	
	In \fref{fig:E} we observe \SI{260}{\giga\pascal} and \SI{270}{\giga\pascal} mean values of the Young's modulus corresponding to  \SI{90}{\degree}-oriented beam and \SI{0}{\degree}, \SI{60}{\degree}, \SI{120}{\degree}-oriented beams, respectively. 
	This result highlights that the AlN film has 3-fold in-plane symmetry, meaning that a \SI{120}{\degree} rotation of the crystal structure results in the same atomic arrangement as before the transformation.
	A \SI{60}{\degree} in-plane rotation of AlN effectively has the same chain of Al-N bonds, resulting in the same elastic properties and therefore the value for Young's modulus at \SI{0}{\degree} and \SI{60}{\degree} orientations. 
	
	\section{Dissipation dilution}
	To derive an expression for the dissipation dilution, we follow the approach that was outlined in Ref. \cite{fedorovMechanicalResonatorsHigh2021}.
	In the theory of elasticity, which describes the mechanics of deformations, the displacement of every point in the membrane is defined through a displacement vector $\Vec{u}$.  
	The definition of strain\footnote{The indices $i,j$ run over $x$, $y$ and $z$.} is
	\begin{equation}
		\epsilon_{ij} = \frac{1}{2}\left(  \underbrace{\frac{\partial u_{i}}{\partial x_{j}} +  \frac{\partial u_{j}}{\partial x_{i}}}_\textrm{linear} + \underbrace{\frac{\partial u_{k}}{\partial x_{i}} \frac{\partial u_{k}}{\partial x_{j}}}_\textrm{nonlinear}  \right),
		\label{eq:nonlinear_strain}
	\end{equation}
	where we have the linear and nonlinear contributions to the strain \cite{fedorovMechanicalResonatorsHigh2021}. In most cases the material enters the plastic regime, before the nonlinear contributions become relevant. However, structures with reduced dimensions, such as nanomechanical resonators with a large aspect ratio, the linear contribution can be equal to zero while the nonlinear contribution is dominant \cite{fedorovMechanicalResonatorsHigh2021}.
	The elastic energy that is stored in such a structure is given as \cite{fedorovMechanicalResonatorsHigh2021}
	\begin{equation}
		w = \frac{1}{2} \sigma_{ij} \epsilon_{ij},
		\label{eq:elastic_energy}
	\end{equation}
	where $\sigma_{ij}$ refers to the stress tensor.
	Since we focus on the dynamics of the system, we separate the static deformation $\bar{x}_{i}(\bf{r})$ and the time-dependent displacement field $u_{i}(\bf{r},t)$ \cite{fedorovMechanicalResonatorsHigh2021}. 
	One can apply this treatment to strain $\epsilon_{ij}$, stress $\sigma_{ij}$ and elastic energy $w$ generated through the acoustic field. We refer to the time-dependent term with $\Delta$ \cite{fedorovMechanicalResonatorsHigh2021} and the strain can be written as
	\begin{equation}
		\epsilon_{ij}(\bf{r},t) = \bar{\epsilon}_{ij}(\bf{r}) + \Delta \epsilon_{ij}(\bf{r},t).
	\end{equation}
	
	\subsection{Anisotropic case} 
	Now we apply a plain stress, \eref{eq:stress_plane}, on a nanomechanical resonator of thickness $h$ with an elasticity matrix of a general form \eref{eq:C_general}.
	Since we are only concerned with flexural membrane modes, the membrane is completely characterized through a slice at $z=0$ and $u(x,y,t) \equiv u_{z}(x,y,0,t)$ \cite{fedorovMechanicalResonatorsHigh2021}. In the neutral plane $(z=0)$ the distances between the points do not change upon small flexural deformations. This constrains the in-plane components of the displacement as
	\begin{equation}
		u_{x}(x,y,z,t) = -z \frac{\partial u(x,y,t)}{\partial x}, u_{y}(x,y,z,t) = -z \frac{\partial u(x,y,t)}{\partial y}.
		\label{eq:deformations}
	\end{equation}
	When we insert \eref{eq:deformations} into \eref{eq:nonlinear_strain}, the general equation for the strain is
	\begin{equation}
		\Delta \epsilon_{ij} = -z \frac{\partial^2 u}{\partial x_{i} \partial x_{j}} + \frac{1}{2}\frac{\partial u}{\partial x_{i}} \frac{\partial u}{\partial x_{j}},
		\label{eq:strain_in_plane}
	\end{equation}
	where we can replace $\epsilon_{ij}$ through its time-dependent part $\Delta \epsilon_{ij}$\cite{fedorovMechanicalResonatorsHigh2021}. When we insert \eref{eq:strain_in_plane} in \eref{eq:elastic_energy}, integrate over the entire mechanical resonator, and take the time-average, we receive the result for $\left<\Delta W^\textrm{(lin)}\right>$ in Eq.~5 in the main text. 
	The non-linear contribution to the elastic energy is \cite{fedorovMechanicalResonatorsHigh2021}
	\begin{equation}
		\left< \Delta W^\textrm{(nl)}\right> = \frac{1}{2} \iint_{V} \bar{\sigma}_{ij} \frac{\partial u}{\partial x_{i}} \frac{\partial u}{\partial x_{j}}dV
		\label{eq:non_linear_elastic_energy}
	\end{equation}
	With these two contributions we can calculate the dissipation dilution factor $D_{Q}$ as \cite{fedorovMechanicalResonatorsHigh2021}
	\begin{equation}
		D_{Q} = 1 + \frac{\left< \Delta W^\textrm{(nl)} \right>}{\left< \Delta W^\textrm{(lin)} \right>}.
	\end{equation}
	
	\subsection{Isotropic case}
	For the isotropic case the stiffness matrix has the form:
	\begin{equation}
		A \cdot \left( \begin{array}{cccccc}
			1-\nu & \nu & \nu & 0 & 0 & 0    \\
			\nu & 1-\nu & \nu & 0 & 0 & 0   \\
			\nu & \nu & 1-\nu & 0 & 0 & 0 \\
			0& 0 & 0 & \frac{1-2\nu}{2} & 0 & 0 \\
			0& 0 & 0 & 0 & \frac{1-2\nu}{2} & 0 \\
			0& 0 & 0 & 0 & 0 & \frac{1-2\nu}{2}
		\end{array}\right),
		\label{eq:isotropic_inplane_stresses}
	\end{equation}
	with $A = \frac{E}{(1+\nu)(1-2\nu)}$.
	
	When we replace the components $C_{ij}$ in Eq.~5 with the components for the isotropic case, we regain the expression for $W^\textrm{(lin)}$\cite{fedorov2019generalized}
	\begin{widetext}
		\begin{eqnarray}
			\Delta W^\textrm{(lin)}_\text{iso}&& = \frac{1}{2}\frac{E h^{3}} {12(1-\nu^2)}\iint\limits_S \left( \frac{\partial^2{u}}{\partial x^2} +  \frac{\partial^2{u}}{\partial y^2}\right)^2 +  (1-\nu) \left(
			\biggl( \frac{\partial^2{u}}{\partial x\partial y} \biggr)^2-2\frac{\partial^2{u}}{\partial x^2}\frac{\partial^2{u}}{\partial y^2} \right)
			dS.
		\end{eqnarray}
		\label{eq:density_bending_energy_isotropic}
	\end{widetext}
	
	\section{Mechanical resonator characterization}
	\subsection{Rounding corners}
\begin{figure}[h!tbp]
		\includegraphics[width = 0.48\textwidth]{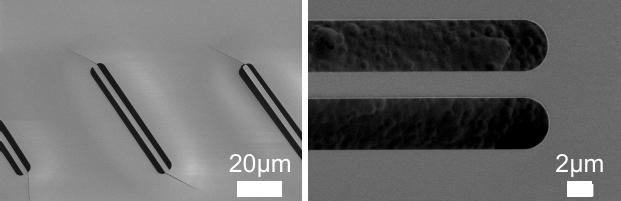}\
		\caption{SEM image of suspended beams without (left, cracks) and with rounding corners (right, no cracks).}
		\label{fig:SEM_crack}
	\end{figure}

	In strained AlN film, we observed crack formation similar to other crystalline material, like InGaP \cite{manjeshwar2023high}. The concentration of stress in sharp corners leads to the formation of cracks along certain crystal directions. 
	AlN shows \SI{60}{\degree}-orientation of the cracks, as seen in \fref{fig:SEM_crack}. To overcome this, it suffices to make at least \SI{2}{\micro\meter} rounding of corners, which we implemented in all geometries.
	
\begin{figure}[h!tbp]
		\includegraphics[width=0.45\textwidth]{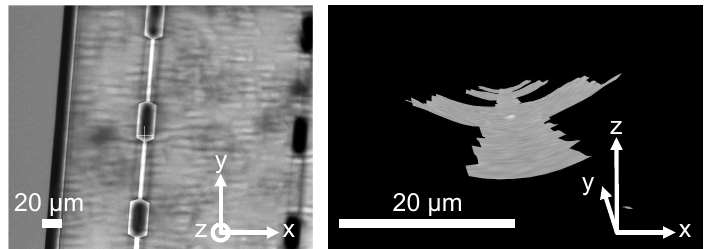}
		\caption{Confocal microscope image of a 1D PnC beam fabricated in AlN.}
		\label{fig:confoc}
	\end{figure}

	AlN has a large residual stress in the film and tends to bend up if the suspended area is not strained, where it can partially relax. For instance, the trapezoid PnC pattern demonstrated strong buckling, see \fref{fig:confoc}, and was prone to cracking if the corners were not rounded. 
	Due to these undesired effects, we opted for elliptically-shaped forms in the unit cells of the 1D PnC beams.
	
\begin{figure}[h!tbp]
		\includegraphics[width=0.45\textwidth]{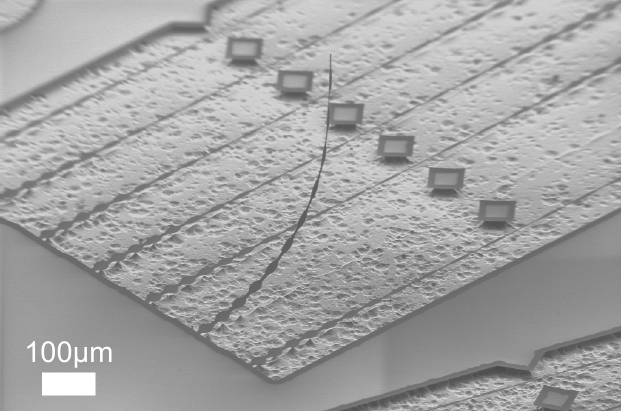}
		\caption{SEM of a \SI{0.7}{\milli\meter}-long broken 1D PnC, showing a strong upward bend.}
		\label{fig:sem_stand}
	\end{figure}

	The bending of free standing structures originates from the inhomogeneous strain in the grown AlN layers. One of the main contributions in such inhomogeneous strain comes from the defects at the interface between silicon and AlN film, which generate a more compressive strain component at the interface upon relaxation \cite{cimalla2007group}. As a result cantilever-like structures bent up \cite{cimalla2007group}, as shown in \fref{fig:sem_stand}.
	
\begin{figure}[h!tbp]
		\includegraphics[width = 0.48\textwidth]{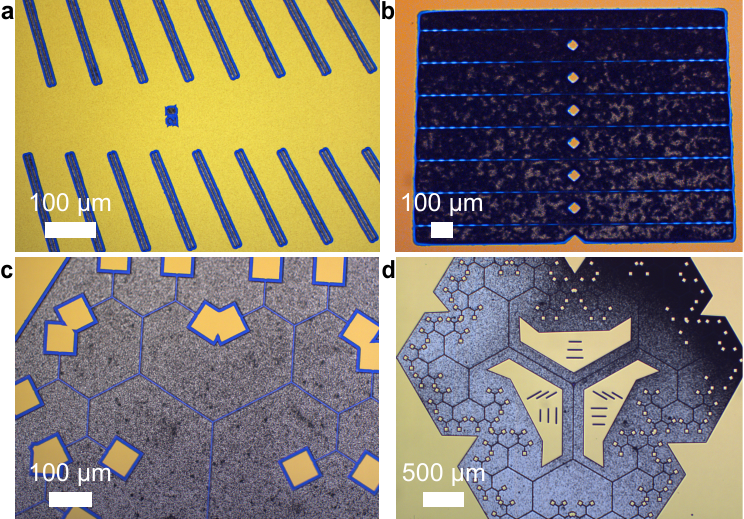}
		\caption{Optical microscope images. a. An array of tensile-strained AlN beams rotated with respect to the AlN crystal. b. An array of tapered 1D PnC beams. c, d. A hierarchically-clamped triangline resonator with $N=6$ branchings and a total tether-length of 4.7\,mm.}\
		\label{fig:uscope}
	\end{figure}

	We additionally show optical microscope images of suspended AlN nanomechanical resonators in \fref{fig:uscope}.
	
	\subsection{Power-dependent measurements}
	
	Thermal effects may influence the measurement of $Q_m$. These would appear as a thermal drift or photothermal effect of the mechanical properties of the resonator. In our case, we would not expect absorption in the bulk of AlN, as the bandgap of AlN is 6.2\,eV and therefore much larger than the photon energy at 1550\,nm. However, as our structures have a defect-rich AlN layer and since absorption can happen at surface defects, we performed additional measurements.
	
\begin{figure*}[h!tbp]
		\centering
		\includegraphics[width=\textwidth]{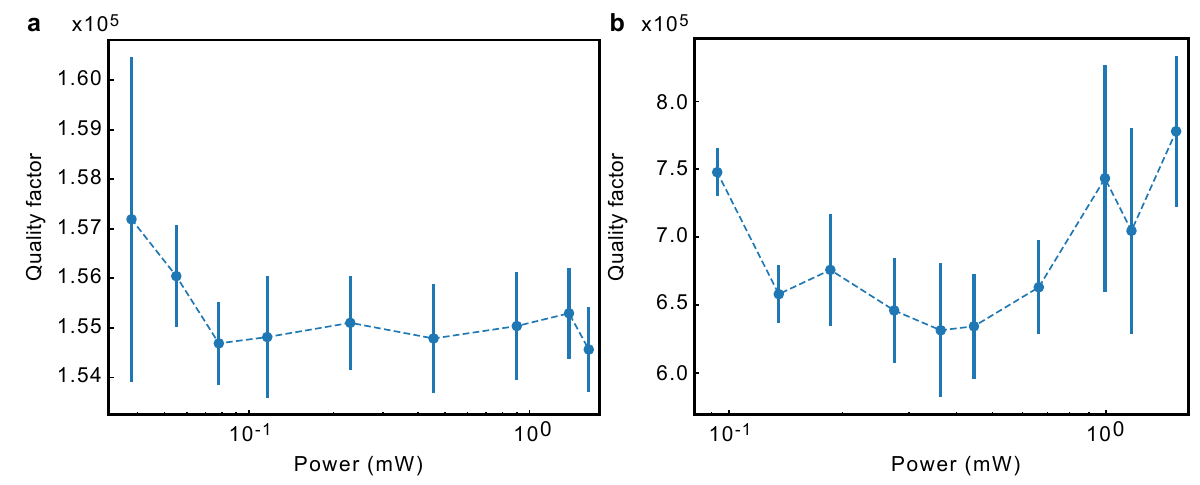}
		\caption{\Qm{} in dependence on the incident laser power used for measuring AlN (a) uniform beam of 75\,$\mu$m length and frequency of \SI{1.57}{\mega \hertz} and (b) 1D PnC with a defect length of $L_{d} = 110\,\mu$m and frequency of \SI{2.504}{\mega \hertz}.
		}
		\label{fig:Q_Pow}
	\end{figure*}

	We measured the mechanical quality factor of an AlN uniform beam and 1D PnC resonators for different incident laser powers, see \fref{fig:Q_Pow}. In this measurement we observed no clear influence on the measured $Q_m$ when changing the impinging optical power by more than an order of magnitude, from 40\,$\mu$W to 1.6\,mW. Hence, we conclude that we do not observe power dependent effects on the mechanical properties in our measurements.
	
	\subsection{1D PnC nanobeams: detailed FEM simulations and additional measurements}
	\begin{figure}[h!tbp]
		\includegraphics[width=0.48\textwidth]{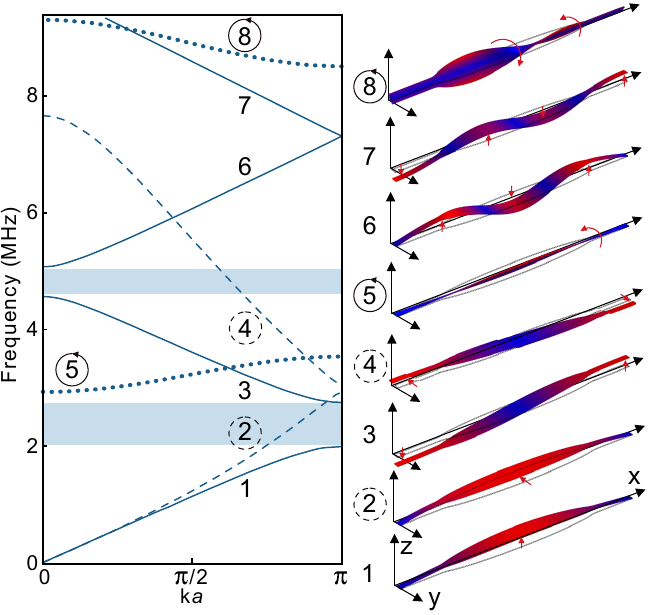}
		\caption{FEM band diagram with the illustrated mechanical modes of the unit cell. We plot the modes that transform under $(P_\textrm{y},P_\textrm{z})$ as $(1,-1)$ as solid lines. The modes that transform as $(-1,1)$ are plotted as dashed lines. The modes that transform as $(-1,-1)$ are plotted as dotted lines.
		}
		\label{fig:banddia}
	\end{figure}
	
	\fref{fig:banddia} shows the band diagram and mode shapes for a unit cell size of \SI{90}{\micro \meter} for $i=0$. We categorize the mechanical displacement field of the different modes with regards to their symmetry under parity operations $P_\textrm{x,y,z}$ \cite{Safavi-Naeini:10}. Here, $x$, $y$ and $z$ refer to the coordinate the parity operation refers to. 
	For instance, mode (1) and (3) transform under the parity operations as $(P_\textrm{y},P_\textrm{z}) = (1,-1)$, while mode (2) transforms as $(P_\textrm{y},P_\textrm{z}) = (-1,1)$.
	Modes that do not transform under the same parity operation do not interact unless the particular symmetry is broken.
	Therefore, the bandgap is created between modes (1) and (3), while mode (2) that crosses it transforms differently under parity transformations.
	
	\begin{figure}[h!tbp]
		\includegraphics[width=0.48\textwidth]{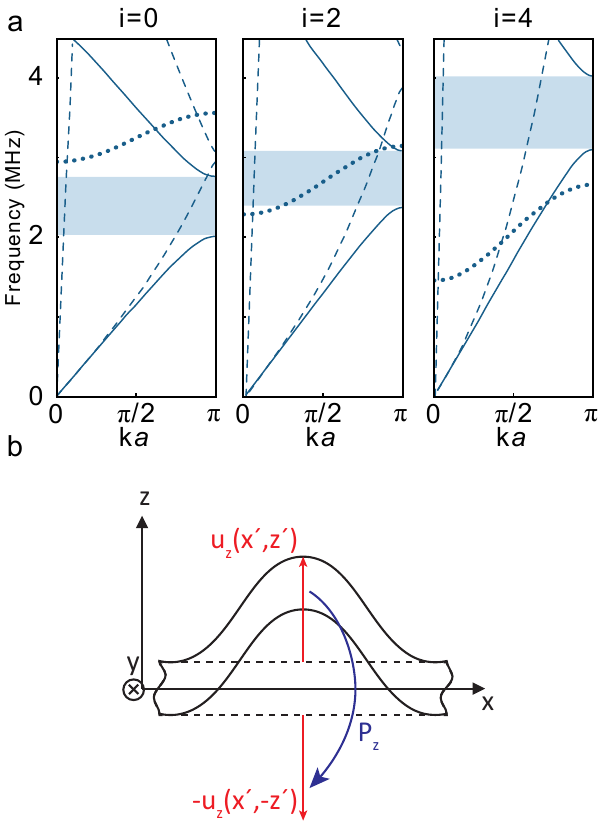}
		\caption{FEM-simulated band diagram for differently sized unit cells, $i = 0, 2, 4$, of the PnC beam. We plot the modes that transform under $(P_\textrm{y},P_\textrm{z})$ as $(1,-1)$ as solid lines. The modes that transform as $(-1,1)$ are plotted as dashed lines. The modes that transform as $(-1,-1)$ are plotted as dotted lines. Schematic drawing of the displacement of the fundamental mode of a beam under a $P_{z}$ parity transformation.}
		\label{fig:band}
	\end{figure}
	
	\fref{fig:band} shows band diagrams for three differently sized unit cells: $i=0$ has a size of \SI{90}{\micro \meter}, $i=2$ has a size of \SI{78.5}{\micro \meter} and $i=4$ has a size of \SI{61.9}{\micro \meter}. We observe that a mode that transforms under the parity transformation $(-1,-1)$ crosses the bandgap. When the symmetry with regards to the $y$-axis is broken (e.g., through fabrication imperfections or buckling of the released devices), this mode could interact with the $(1,-1)$ modes and reduce soft clamping, e.g., in the unit cell $i=2$.
	
	\begin{figure}[h!tbp]
		\includegraphics[width=0.48\textwidth]{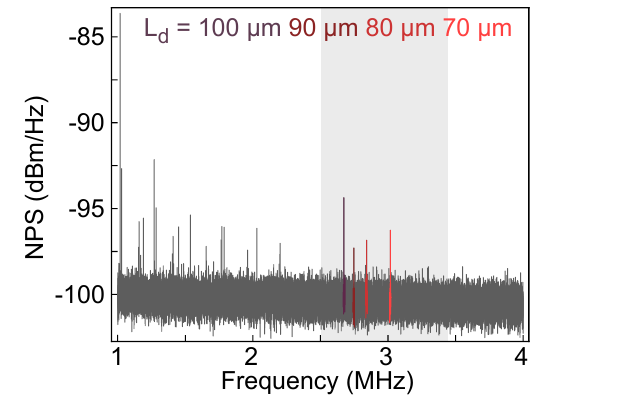}
		\caption{NPS spectra of four 1D PnC beams with different defect lengths, $L_d$.}
		\label{fig:NPS_Defect}
	\end{figure}
	
	In \fref{fig:NPS_Defect} we display additional NPS spectra of four PnC beams with different defect lengths ($L_d = $70, 80, 90, \SI{100}{\micro\meter}). As expected, the shorter the defect length, the larger the eigenfrequency of the defect mode. With the sweep of $L_d$ we obtain the defect mode crossing through the bandgap in Fig.2 in the main text.
	
	\subsection{Singly-branched nanobeams}
	\begin{figure}[h!tbp]
		\includegraphics[width = 0.48\textwidth]{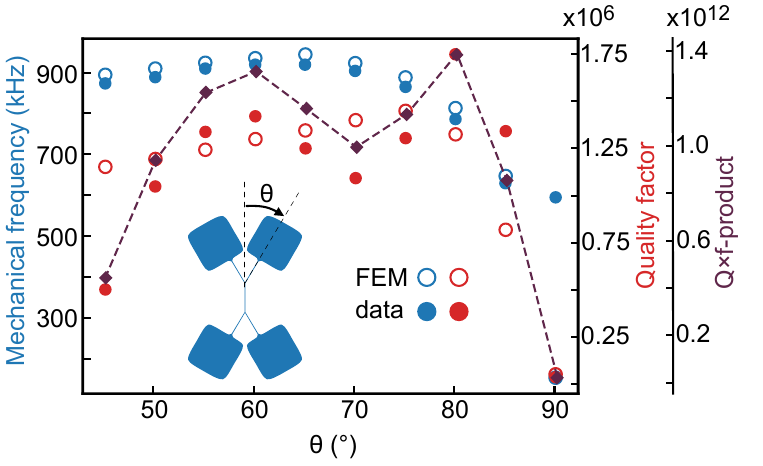}
		\caption{\fm{} (blue) and \Qm{} (red) of singly-branched resonators in dependence on the branching angle, $\theta$. Experimental data are dots and simulations are open circles. Purple rhombus shows experimental values for the \Qf-product.}
		\label{fig:branch}
	\end{figure}
	
	We fabricated \SI{300}{\micro \meter}-long beams with one iteration of branching and investigated the \fm{} and \Qm{} dependence on branching angle, $\theta$, see \fref{fig:branch}. As seen in \fref{fig:branch}, the experimental values of \fm{} follow closely FEM predictions with a maximum frequency around $\theta = $ \SI{60}{\degree}.
	While the FEM simulation predicts a maximum $Q_{D}$ around \SI{75}{\degree}, the experimental data shows a local maximum at $\theta = $ \SI{60}{\degree} and a global maximum at \SI{80}{\degree}. 
	The local maximum could be the result of the \SI{60}{\degree}-periodicity of \sr{}, which produces equal tension in the branched beams at $\theta = $ \SI{60}{\degree}. The global maximum at $\theta =$ \SI{80}{\degree} is expected according to the results of isotropic case \cite{fedorov2019generalized}.
	We also plot the experimental \Qf-product in \fref{fig:branch}. We observe that the \Qf{}-product peaks at $\theta = $ \SI{60}{\degree} and \SI{80}{\degree}, where the latter yields a slightly higher value.
	
	We choose a branching angle of $\theta = $ \SI{60}{\degree} for fabricating the hierarchically-clamped traingline structures. This branching angle results in beam orientations that follow the \SI{60}{\degree}-periodicity in $E$, \sr{}, at the same time it yields a large \Qf{} value.
	
	\subsection{Hierarchically-clamped triangline vs.~trampoline resonators}
	\begin{figure}[h!tbp]\
		\includegraphics{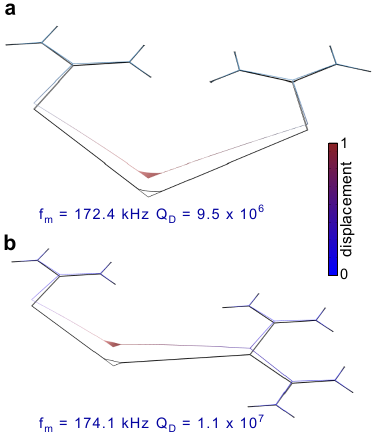}\
		\caption{FEM simulations of the fundamental mode for a hierarchically-clamped trampoline (a) and triangline (b).}\
		\label{fig:comparison}
	\end{figure}\
	
	We compare different trampoline-like geometries for the same central pad size. 
	Specifically, we simulate a conventional trampoline (4 tethers) and the triangline (3 tethers) with the same prestress of the film (\sg{} = \SI{1.4}{\giga\pascal}), branching angle ($\theta =$ \SI{60}{\degree}), number of branching iterations ($N = 3$), constant width of the tether, and total tether-length of \SI{830}{\micro \meter}. We show the fundamental mode of the FEM simulations in \fref{fig:comparison}.
	For this comparison we removed the PhC from the central pad. The mechanical frequency of the resonators varies only slightly in the FEM simulation, the trampoline's \fm{} = \SI{172.4}{\kilo\hertz} and triangline's \fm{} = \SI{174.1}{\kilo\hertz}. The mechanical quality factor is slightly higher for the triangline geometry: triangline has $Q_D = 1.06\times 10^7$ and the trampoline has $Q_D = 9.54\times 10^6$.
	
	Besides that the triangline with a branching angle of $60^\circ$ follows the in-plane crystal structure, it has the additional advantage of allowing to increase the number of branching iterations.
	
	\subsection{Piezoelectricity of AlN thin films and mechanical frequency tuning}
	
	\subsubsection{Piezoelectricity of AlN thin films}

	We summarize relevant works that determined the piezoelectic coefficient in thin-film AlN in Tab.\ref{tab:piezo}. The bulk single-crystal piezoelectric coefficients are $d_{33} = \SI{5.6}{\pico \meter/ \volt}$ and $d_{31}  = -d_{33}/2$ \cite{guy1999extensional}, while polycrystalline AlN has a $d_{33}$ between \SI{3.4}{\pico \meter/ \volt} and \SI{5.15}{\pico \meter/ \volt} \cite{fei2018aln}.
	
	\begin{table}[h!tbp]
		\centering
		\begin{tabular}{c|c|c|c}
			Growth method & thickness (nm) & $d_{33}$ (pm/V) & Ref \\
			\hline        
			DC-sputtering & 1000 & 3.4 & \cite{dubois1999properties}\\
			DC-sputtering & 50 & 3.51 & \cite{howell2019effect}\\
			CVD & 1000 & 4.0 & \cite{guy1999extensional}\\
			MOCVD & 130-250 & 5.47 & 
			\cite{tonisch2006piezoelectric}\\
			Bulk &  & 5.6 & \cite{guy1999extensional}\\
		\end{tabular}
		\caption{Piezoelectric coefficient of AlN thin films grown with different methods and compared to the bulk value.}
		\label{tab:piezo}
	\end{table}
	
	The highest piezoelectric performance would be achieved by a defect-free, epitaxial film with orientation along the $c$-axis \cite{tonisch2006piezoelectric}. In non-epitaxially grown films, the varying orientation of the domains in the polycrystalline film may result in reduced piezoelectricity.

	\subsubsection{Piezoresponse force
		microscopy: piezoelectric constant of our AlN thin film}
	
	As we showed in Section \ref{sec:xrd}, the FWHM of the (0002) reflex is \SI{0.3}{\degree} in our MOVPE-grown AlN thin films, while a sputtered polycrystalline AlN film with $d_{33} =\SI{3.51}{\pico \meter/ \volt}$ has a FWHM of the (0002) reflex above \SI{2}{\degree} \cite{howell2019effect}. Our small FWHM value therefore indicates that our film has a high crystalline quality and we therefore expect the film to exhibit piezoelectric behavior.
	
	We performed piezoresponse force microscopy (PFM) \cite{tonisch2006piezoelectric, soergel2011piezoresponse} with a Bruker Dimension ICON AFM to determine the piezoelectric constant of our AlN film. 
	For that measurement we utilize the inverse piezoelectric effect, where an AC modulation voltage is applied between the AFM tip and a conducting substrate, which in our case is a doped Si wafer that has a resistivity of $0.1\,\Omega\cdot$cm. This causes the piezoelectric film on top of the substrate to expand and contract at the same frequency as the applied voltage. The induced film displacement is then measured via the displacement of the AFM tip.
	
\begin{figure}[h!tbp]
		\centering
		\includegraphics[width=0.45\textwidth]{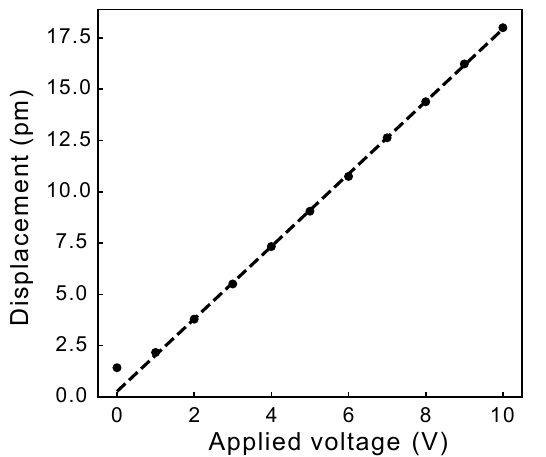}
		\caption{Measured mechanical displacement vs. applied voltage. The amplitude of the AC component of the applied voltage was varied from 0 to 10\,V at a drive frequency of 50\,kHz. Dashed line is a linear fit to extract the piezoelectric coefficient and we obtain 1.8\,pm/V.}
		\label{fig:piezo}
	\end{figure}

	\fref{fig:piezo} shows the vertical displacement of the AFM tip versus the applied AC voltage. We observe a linear increase of displacement with an increase in voltage, as expected from a linear piezoelectric response of the AlN thin film. Note that we performed the measurement on a non-suspended part of the AlN film as the AFM tip is required to be in contact mode with the surface for this PFM measurement. We also measured the induced displacement in a scan window of 10\,$\mu$m$\times$10\,$\mu$m and obtained homogeneous displacement with no domains observable. We verified that we do not observe any piezoelectrically-induced displacement when measuring on a part of the Si wafer, where the AlN layer was not present. Hence, we conclude that the measured PFM signal originates from the piezoelectric response of the AlN film.
	
	The slope of the curve in \fref{fig:piezo} yields the effective piezoelectric coefficient $d_{33,\text{eff}}=1.8 \pm 0.1$\,pm/V of the AlN film. 
	As the AlN film is rigidly clamped to the Si substrate, contraction and expansion of the film are constrained leading to a decrease of its piezoelectric response \cite{torah2004experimental}.
	Further reduction of the piezoelectric coefficient can be related to the \SI{20}{\nano \meter}-thick defect-rich AlN layer at the interface that reduces the total piezoelectric response of the film due to its crystalline disorder, or the large stress in our AlN film that we exploit for realizing high-$Q_m$ nanomechanical resonators.
	
	\subsubsection{Piezoelectric tuning of mechanical frequency}
	
	We estimate the effect of the piezoelectricity onto the eigenfrequency of a uniform AlN beam. For that we take as an example a beam of length $50\,\mu$m and residual stress of 1.4\,GPa. In the absence of an external electric field this beam has a fundamental frequency $f_0 = 1/(2L)\sqrt{\sigma_{11}/\rho} = 5.53$\,MHz. 
	We assume that we apply an electric field along the $c$-axis, i.e., $\vec{E} = (0, 0, E_3)^T$.
	For the hexagonal crystal lattice, the piezoelectric polarization is given by the tensor
	\begin{equation}
		[e]_{3 \times 6} = 
		\begin{bmatrix}
			0 & 0 & 0 & e_{14} & e_{15} & 0\\
			0 & 0 & 0 & e_{15} & -e_{14} & 0\\
			e_{31} & e_{31} & e_{33} & 0 & 0 & 0\\
		\end{bmatrix},
	\end{equation}
	where $e_{33} = 1.55$ C/m$^2$, $\epsilon_{31} = \epsilon_{32} = -0.58$ C/m$^2$, $e_{15} = -0.48$ C/m$^2$, and $\epsilon_{24} = \epsilon_{15} = -0.48$ C/m$^2$ (which corresponds to $d_{33} = 5$\,pm/V).
	The induced stress along the direction of the beam can be calculated as the first component of $\epsilon^\text{T}\cdot E$.
	As $E_3 = V/h$ with the applied voltage $V$ over the thickness $h$ of the resonator, the piezo-induced stress is
	\begin{equation}
		\sigma_{11}^p = \sigma_{11} - e_{31} E_3 = \sigma_{11} - \frac{e_{31} V}{h} = \sigma_{11}^0 - 2[\text{MPa/V}]\times V.
	\end{equation}
	If we apply 5\,V then the shifted mechanical frequency is $f_p =  5.57$ MHz, corresponding to a tuning coefficient of $\Delta f/V = 5.5$ kHz/V.
	
	We can compare this analytical result to simulations that we perform via the multiphysics Comsol model (coupling the solid state and electrostatics models). To this end, we model a 290\,nm-thick AlN layer and apply the E-field along the $c$-axis. The results of the FEM simulation for a beam, 1D PnC and short triangline are shown in Tab.~\ref{tab:FEM_piezo}. We find a good agreement between the FEM results and the simple analytical model. These results show that it is possible to tune the mechanical frequency via applying a DC voltage with a rate of kHz/V. The results also demonstrate that the tunability is strongly dependent on the exact geometry of the nanomechanical resonator.
	
	\begin{table}[h!]
		\centering
		\begin{tabular}{c|c|c|c}
			& \shortstack{$f_m$ (MHz) \\ at 0\,V} & \shortstack{$f_m$ (MHz) \\ at 5\,V} & \shortstack{$\Delta f_m$/V \\ (kHz/V) }\\
			\hline
			uniform beam & 5.79 & 5.83 & 8\\
			1D PnC & 3.39 & 3.41 & 4\\
			triangline & 0.1684 & 0.1694 & 0.2\\
		\end{tabular}
		\caption{FEM results for the mechanical frequency tuning of the 50$\mu$m-long beam, defect mode of the 1D PnC and the fundamental mode of the short triangline. All structures are 290\,nm-thick and have residual stress of 1.4\,GPa.}
		\label{tab:FEM_piezo}
	\end{table}
	
	\subsection{Optical reflectivity measurements}
	\begin{figure}[h!tbp]
		\includegraphics[width = 0.48\textwidth]{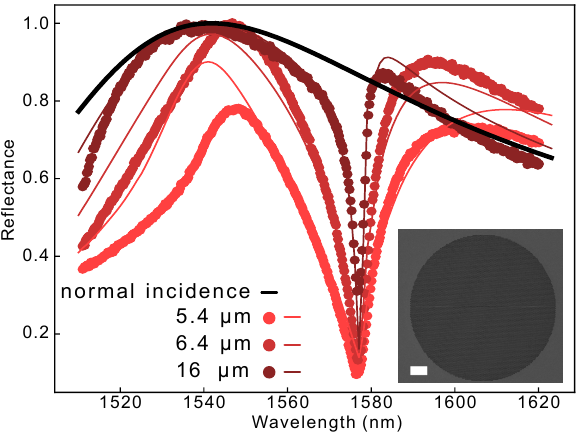}
		\caption{Reflectance of a hexagonal PhC patterned on a circular membrane. The inset shows an SEM image of a \SI{180}{\micro\meter} diameter circular membrane, scale bar \SI{20}{\micro\meter}. Experimental data represented by dots and RCWA simulated reflectance is shown as solid lines. Different colors represent different beam waists. The black solid line is the simulated reflectance at normal incidence of a plane wave for $a_\text{PhC} = \SI{1450}{\nano\meter}$ and $r_\text{PhC} = \SI{508}{\nano\meter}$.}
		\label{fig:R_hPhC}
	\end{figure}
	
	The central pad of the triangline is patterned with a hexagonal PhC. The same hexagonal PhC pattern was used for the \SI{180}{\micro\meter}-diameter circular membrane, \fref{fig:R_hPhC}. 
	We varied the waist of the incident beam from 5 to \SI{16}{\micro\meter} and measured the reflectance of the patterned membrane. A lower reflectance is observed with a smaller beam waist due to higher oblique angles of incidence \cite{feng1996off,manjeshwar2023high}.
	In the main text, the reflectivity measurements were performed with a \SI{6.4}{\micro\meter} waist of a wavelength-tunable \SI{1550}{\nano\meter} laser.
	
\begin{figure}[h!tbp]
		\centering
		\includegraphics[width = 0.45\textwidth]{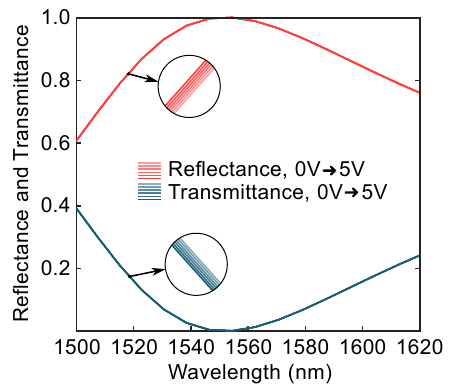}\
		\caption{FEM simulation of the AlN PhC reflectance in dependence on the applied voltage.}\
		\label{fig:Pockels}
	\end{figure}

	The wavelength-dependence of the reflectance can be tuned through the electro-optic effect in AlN. To evaluate this effect, we consider a simplified model of (i) an optical beam at normal incidence to (ii) an infinite AlN photonic crystal slab, where (iii) the AlN is assumed to be an isotropic material with an average electro-optic coefficient of 1\,pm/V \cite{liu2023aluminum}. We simulated the expected reflectivity of a photonic crystal when different voltages are applied, see \fref{fig:Pockels}. We observe that at 1510\,nm the reflectance tuning is about -4.8\,pm/V, whereas at 1600\,nm it is about -9.8\,pm/V. So, the effect of applying a voltage on the PhC reflectance is to shift the overall reflectance curve by some picometers per volt. Further, the unequal tuning at the different sides of the maximum of the reflectance leads to a slight linewidth change of the Fano resonance.
	
	\
	
	\subsection{Trianglines orientation and filling with a PhC}
	\begin{figure}[h!tbp]\
		\includegraphics[width = 0.48\textwidth]{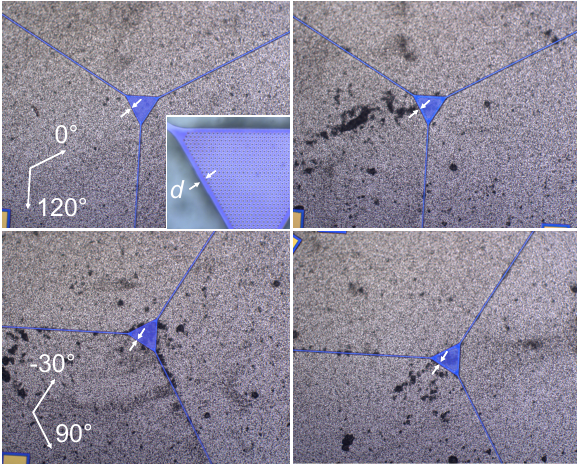}\
		\caption{Optical image of the triangline with varied in-plane orientation. The arrows in the corners indicate the in-plane orientation of the tethers. The arrows on the PhC pad indicate the distance of the PhC pattern to the edge of the pad.}\
		\label{fig:tria_var}
	\end{figure}\
	
	We varied the in-plane orientation of the tethers of the triangline, $\alpha$ (\fref{fig:tria_var}), to analyze how weak in-plane anisotropy of the elastic properties affect \fm{} and \Qm{}. We find no major difference in \fm{} nor \Qm{}, see Tab.~\ref{tab:var_tria}.
	Furthermore, we varied the PhC area on the central \SI{60}{\micro\meter}-side-long pad, such that the distance of the PhC pattern to the edge, $d$, changes from \SI{1.9}{\micro\meter} to \SI{3.4}{\micro\meter}. As can be seen in Tab.~\ref{tab:var_tria}, in case of $d = \SI{3.4}{\micro\meter}$, the triangline is stiffer and exhibits a \SI{4}{\kilo\hertz} higher frequency. 
	
	\begin{table}[h!tbp]
		\centering%
		\begin{tabular}{c| c| c| c}
			\hline
			\hline
			$\alpha$ (\SI{}{\degree}) & $d$ (\SI{}{\micro\meter}) & \fm{} (\SI{}{\kilo\hertz}) & \Qm{} \\
			\hline
			0 & 1.9 & 200.2 & $8.6\times 10^6$\\
			-30 & 1.9 & 200.7 & $9.4\times 10^6$\\
			\hline
			0 & 3.4 & 204.5 & $7.4\times 10^6$\\
			-30 & 3.4 & 204.9 & $8.4\times 10^6$\\
			\hline
			\hline
		\end{tabular}
		\caption{Influence of triangline's in-plane orientation, $\alpha$, and PhC filling of the pad, $d$, on \fm{} and \Qm{}.}
		\label{tab:var_tria}
	\end{table}
	
	\subsection{Gas damping}
	\begin{figure}[h!tbp]
		\includegraphics[width=0.4\textwidth]{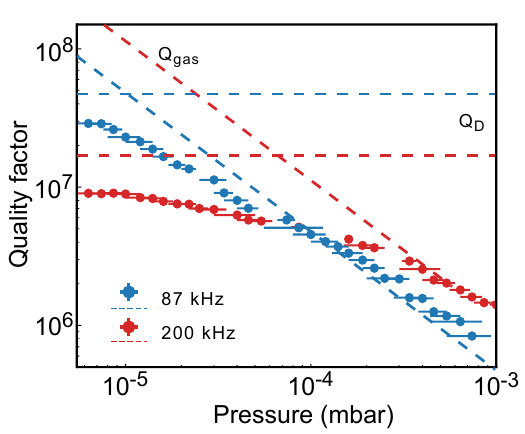}
		\caption{Quality factor vs.~pressure of the short triangline (\SI{200}{\kilo\hertz}) and long triangline (\SI{87}{\kilo\hertz}) fundamental mode. Dashed lines are limiting mechanisms: gas damping, $Q_\textrm{gas}$, and diluted intrinsic loss, \Qd{}.}
		\label{fig:Q_P}
	\end{figure}
	
	\fref{fig:Q_P} shows the pressure dependence of the fundamental mode of trianglines. We observe that for pressures larger than $10^{-4}\,\SI{}{\milli\bar}$ the quality factor is pressure dependent and follows the prediction by viscous gas damping \cite{verbridge2008size}:
	\begin{equation}
		Q^{-1}_\textrm{gas} = \left( \frac{2}{\pi} \right)^{3/2} \frac{P}{\rho h f_\text{m}}\sqrt{\frac{M}{RT}},
		\label{eq:q_gas}
	\end{equation}
	where $M$ is the molecular mass of the gas molecules, $R$ is the molar gas constant, and $T$ is the temperature of the gas. However, at pressures below $10^{-5}\,\SI{}{\milli\bar}$, gas damping is not the limiting mechanism and we expect that the \Qm{} is limited by dissipation dilution.
	
	\subsection{Nanomechanical polygon resonators}
	
	A geometry to achieve high \Qm{} is a polygon-type resonator from Ref.~\cite{bereyhi2022perimeter}. For instance, for a six-sided polygon resonator the dilution factor of a sinusoidal standing wave in the perimeter beams of equal stress along all segments is given as \cite{bereyhi2022perimeter}
	\begin{equation}
		D_Q^{-1} = \left( \frac{1}{n^2 \pi^2 \lambda^2}\right)^{-1} + \left( \frac{r_l (1+\nu) \cos{^2(\pi / 6)}}{4r_w  \lambda^2 }\right)^{-1},
	\end{equation}
	where $r_l$ is the ratio of the support length to the side-length of the polygon, $r_w$ is the ratio of the support width to the side-width of the polygon, $n$ is the perimeter mode order, and $\nu$ is the Poisson ratio.
	
	The stress parameter of a uniform beam of length $l_0 = \SI{2}{\milli\meter}$ and thickness \SI{290}{\nano\meter} is $\lambda = 6.8\times10^{-4}$. Then the fundamental perimeter mode of a hexagonal polygon with $r_l = 0.4$, $r_w = 1$ reaches $D_Q = 1.1 \times 10^{5}$ and \Qm{} of $10^9$.
	
	The same hexagon geometry can reach a \Qm{} of $10^{10}$ for a \SI{100}{\nano\meter}-thick AlN layer assuming that the material properties stay the same.
	With the eigenfrequency of the perimeter mode of \SI{138}{\kilo\hertz} we would obtain a \Qf-product of $10^{15}$\,Hz.

%

	\clearpage

\end{document}